\documentclass[pra,%
 aps,twocolumn,%
 amsmath,amssymb,
 superscriptaddress,
 reprint,%
]{revtex4-2}

\usepackage{graphicx}
\usepackage{dcolumn}
\usepackage{bm}

\usepackage{etoolbox}
\usepackage{braket}
\usepackage{hyperref}
\usepackage{algorithm}
\usepackage{algpseudocode}
\usepackage{setspace}
\usepackage{float}
\usepackage{placeins}
\usepackage{multirow}
\usepackage{booktabs}
\usepackage{nicematrix}
\usepackage{todonotes}
\usepackage{physics}
\usepackage[normalem]{ulem}
\usepackage{bbm}

\definecolor{forestgreen(web)}{rgb}{0.13, 0.55, 0.13}

\algnewcommand{\algorithmicor}{\textbf{ or }}
\algnewcommand{\continue}{\textbf{continue}}
\algnewcommand{\OR}{\algorithmicor}

\DeclareMathOperator*{\argmin}{arg\,min}
\DeclareMathOperator*{\argmax}{arg\,max}

\begin{document}

\preprint{AIP/123-QED}

\title{Problem specific ion native ansatz for combinatorial optimization}

\author{Georgii Paradezhenko}%
 \email{g.paradezhenko@skoltech.ru}
 \affiliation{Skolkovo Institute of Science and Technology, Moscow, Russian Federation}%
\author{Daniil Rabinovich}%
\affiliation{The Russian Quantum Center, Moscow, Russian Federation}%
\affiliation{Skolkovo Institute of Science and Technology, Moscow, Russian Federation}%
\affiliation{Moscow Institute of Physics and Technology, Dolgoprudny, Russian Federation}%
\author{Ernesto Campos}%
\affiliation{Skolkovo Institute of Science and Technology, Moscow, Russian Federation}%
\author{Kirill Lakhmanskiy}%
\affiliation{The Russian Quantum Center, Moscow, Russian Federation}%

\date{\today}

\begin{abstract}

Variational quantum algorithms have become a standard approach for solving a wide range of problems on near-term quantum computers. Identifying an appropriate ansatz configuration for variational algorithms, however, remains a challenging task, especially when taking into account restrictions imposed by real quantum platforms. This motivated the development of digital-analog quantum circuits, where sequences of quantum gates are alternated with natural Hamiltonian evolutions. 
A prominent example is the use of the controllable long-range Ising interaction induced in ion-based quantum computers. This interaction has recently been applied to develop an algorithm similar to the quantum approximate optimization algorithm (QAOA), but native to the ion hardware. 
The performance of this algorithm has demonstrated a strong dependence on the strengths of the individual ion-ion interactions, which serve as ansatz hyperparameters. 
In this work, we propose a heuristic for identifying a problem-specific ansatz configuration, which enhances the trainability of the ion native digital-analog circuit. The proposed approach is systematically applied to random instances of the Sherrington-Kirkpatrick Hamiltonian for up to 15 qubits, providing favorable cost landscapes. As a result, the developed approach identifies a well-trainable ion native ansatz, which requires a lower circuit depth to solve specific problems as compared to standard QAOA. This brings the algorithm one step closer to its large scale practical implementation.

\end{abstract}

\maketitle

\section{\label{sec:intro}Introduction}


Current day quantum computers are inherently limited by both environmental and systematic noise. In the absence of error correction, these devices can only support the execution of short depth quantum circuits which fit within their coherence times~\cite{weidenfeller2022scaling,ratcliffe2018scaling,hegde2022toward, mills2022two}. 
Within these restrictions, the variational model of quantum computing has been developed to explore the full potential of current hardware. In a variational quantum algorithm (VQA), a short depth parametrized quantum circuit is iteratively optimized by a classical co-processor in an attempt to minimize a given cost function \cite{cerezo2021variational}. This model has been proven to be universal \cite{biamonte2021universal} adapted for a wide variety of tasks, notably: the variational quantum eigensolvers (VQEs) \cite{parrish2019quantum,hempel2018quantum,kandala2017hardware, peruzzo2014variational, cade2020strategies,zeng2021simulating, kardashin2021numerical, uvarov2020frustrated, uvarov2024mitigating} designed to approximate the ground energy of Hamiltonians, Hamiltonian simulations \cite{yuan2019theory,cirstoiu2020variational,gibbs2021long,endo2020variational}, the quantum approximate optimization algorithm (QAOA) \cite{Farhi2014,morales2020universality, farhi2014quantumBOCP,wang2018quantum,lloyd2018quantum,zhou2020quantum,claes2021instance,wauters2020polynomial,rabinovich2022progress,akshay2021parameter,akshay2020reachability,campos2024depth,campos2021training} designed to give approximate solutions to combinatorial optimization problems, quantum circuit compilers \cite{khatri2019quantum,jones2022robust, he2021variational},
machine learning \cite{liang2020variational,mcclean2016theory, benedetti2019parameterized,verdon2017quantum}, among others \cite{mitarai2018quantum,schuld2019evaluating,xu2019variational,cerezo2021cost}.

Despite the benefits offered by the variational model \cite{gentini2020noise, 
sharma2020noise, cincio2021machine, harrigan2021quantum,guerreschi2019qaoa,pagano2020quantum}, inaccurate results arising from gate imperfections and restricted circuit depth remain a prime concern~\cite{rabinovich2024robustness,akshay2022circuit}. 
One pathway towards alleviating these limitations is the use of digital-analog quantum circuits which alternate between quantum gates and entangling natural Hamiltonian evolutions \cite{yu2022superconducting, garcia2024digital, martin2020digital, gonzalez2021digital}, reducing the circuit dependence on entangling gates of limited fidelity. A typical example is the use of the long-range Ising interaction in trapped ion-based quantum computers \cite{zhang2017observation, maier2019environment, monroe2021programmable, richerme2014non, smith2016many, senko2014coherent}, which has been applied to quantum Hamiltonian simulations \cite{parra2020digital}, QAOA compilation \cite{pagano2020quantum, headley2022approximating} and used as an entangling layer in VQE circuits~\cite{zhuang2024hardware, kokail2019self}.
In the same fashion, \cite{rabinovich2022ion} has proposed to use this interaction as a substitute to the problem Hamiltonian exponent to generate a QAOA-like sequence capable of minimizing general $\mathbb{Z}_2$ symmetric combinatorial problems. The performance of this approach strongly depends on the choice of ion interactions referred to as ansatz hyperparameters.
Indeed, problem agnostic hyperparameters can lead to trainability limitations induced by a large circuit expressivity~\cite{holmes2022connecting} or by cost landscapes plagued with local minima \cite{AK2022}. On the other hand, as demonstrated in \cite{rabinovich2022ion}, certain problem-specific ion native ansatz hyperparameters can significantly improve the algorithmic performance, yet a procedure to identify them has been lacking so far.

Following the framework in \cite{rabinovich2022ion}, we develop an approach for identifying hyperparameters suitable for minimizing  specific problem instances. For a QAOA-like ansatz we propose a heuristic based on optimizing a single layer circuit expectation value alternately over variational and circuit hyper- parameters using the block coordinate descent method (see, e.g.,~\cite{Tseng2001}).
The proposed heuristic is benchmarked by performing Hamiltonian minimization for the Sherrington-Kirkpatrick (SK) model from $5$ to $15$ qubits with the coupling coefficients sampled from the standard normal distribution. This is a widely studied model in condensed matter physics, arising as a mean field approach to spin glasses~\cite{panchenko2012sherrington,sherrington1975solvable}. 
We demonstrate that the proposed heuristic identifies an ansatz structure, which provides a favorable optimization landscape. When compared to standard QAOA~\cite{Farhi2014} and the previous approach proposed in \cite{rabinovich2022ion}, we establish that our heuristic ameliorates the optimization and notably reduces the circuit depth required to solve a problem. We argue that this behavior is associated to ansatz states getting effectively locked to a low-dimensional subspace that has support on the low energy states of the problem Hamiltonian. 

The paper is structured as follows. Sec.~\ref{sec:QAOA} provides an outline of the ion native digital-analog circuit implementation based on the long-range Ising interactions. In Sec.~\ref{sec:heuristics}, we introduce a heuristic for searching problem-specific hyperparameters of the ion native ansatz that ameliorate the optimization in VQA. Sec.~\ref{sec:results} shows the numerical results, where we benchmark the proposed heuristic on random instances of the SK model and investigate how the choice of hyperparameters affects the ion-based QAOA performance. Sec.~\ref{sec:conclusions} summarizes and discusses the results.




\vspace{-0.5em}
\section{\label{sec:QAOA}Ion native quantum ansatz}
\subsection{Global ion interaction}
Trapped ion-based quantum computers are mostly known for their fully connected architecture, which allows executing entangling gates between any pair of ions in a trap \cite{pogorelov2021compact, wright2019benchmarking}. 
The interaction between ions is turned by the excitation of collective ion oscillation modes (i.e. phonons) at specific frequencies $\omega_m$ under laser radiation. If, however, the driving frequency is tuned far from the phonon frequencies (dispersive regime), the phonons are only virtually excited, inducing the evolution under  the effective Hamiltonian~\cite{ZMD2006,monroe2021programmable} (for details, see Appendix~\ref{appendix:ions}),
\begin{equation}\label{HI-native}
    H_{\mathrm{I}} =\sum_{i < j} J_{ij} X_i X_j,
\end{equation}
with $X_j$ being Pauli $X$ operator applied to the $j$-th qubit. The coupling coefficients in Eq.~\eqref{HI-native} are given by
\begin{equation}\label{J-def}
    J_{ij} =
    \Omega_i \Omega_j 
    \sum_m \frac{\eta_i^m \eta_j^m \omega_m}{\mu^2 - \omega_m^2}
    \equiv \Omega_i \Omega_j C_{ij}.
\end{equation}
Here, the matrix $C_{ij}$ depends on the spectrum of phonon frequences $\omega_m$, Lamb-Dicke parameters $\eta_j^m$ which quantify the displacement of the $j$-th ion in the $m$-th phonon mode, and laser detuning $\mu$ from the carrier frequency. The matrix $C_{ij}$ thus can be precalculated and kept fixed (for details, see Appendix~\ref{appendix:phonons}). 
A characteristic feature of the interaction~\eqref{J-def} is that it can be controlled in the experiment by varying the Rabi frequencies $\Omega_i$ induced by a laser field individually for each ion \cite{monroe2021programmable}.
Introducing the parameters $A_j = \Omega_j/\Omega_{\max}$, where $\Omega_{\max}$ is the maximum allowed Rabi frequency in the experiment, the Ising coupling coefficients \eqref{J-def}  can be rewritten as
\begin{equation}\label{J-modified}
    J_{ij} =
    A_i A_j \Omega_{\max}^2 C_{ij},
\end{equation}
where the dimensionless hyperparameters $A_j \in [-1,1]$ control the strength of pairwise interactions. As such, the ion-ion interaction can be characterized by a vector of hyperparameters $\bm{A} = (A_1,\ldots,A_n)$.
\vspace{-1em}
\subsection{Ion native quantum circuit}

Utilizing global interactions in variational circuits is beneficial, as they can naturally entangle the system allowing to avoid the extensive use of entangling quantum logical gates. For the uniform choice of the frequencies $\Omega_j = \Omega$, when the couplings~\eqref{J-def} are approximated as a power decay law $J_{ij} \propto 1/|i-j|^a$, the interaction~\eqref{HI-native} has been used as an entangling block in hardware efficient ansatzes and the QAOA circuit \cite{pagano2020quantum, zhuang2024hardware, kokail2019self}. 
In \cite{rabinovich2022ion} the effective Hamiltonian \eqref{HI-native} with the non-uniform choice of $\Omega_j$ has been used to generate an ion native ansatz in the form of a QAOA-like sequence
\begin{eqnarray}
    \ket{\psi_p(\bm{\beta},\bm{\gamma})} 
    =  \prod_{k=1}^p & & \biggl[ e^{-i\beta_k H_x  }\mathrm{H}_+ e^{-i \gamma_k H_{\mathrm{I}}} \mathrm{H}_+ \biggr] \ket{+}^{\otimes n},
    \label{ion-native-ansatz}
\end{eqnarray}
where $\mathrm{H}_{+} = (\vert + \rangle \langle 0 \vert + \vert - \rangle \langle 1 \vert)^{\otimes n}$ is the Hadamard gate
applied to all the qubits, $H_x = \sum_i X_i$ is the mixer Hamiltonian, and $(\bm{\beta},\bm{\gamma}) \in [0,\pi)^{p} \times [0,2\pi)^{p}$ are the variational parameters.
Similar to the settings of standard QAOA, the ansatz \eqref{ion-native-ansatz} has been applied to solving combinatorial problems~\cite{Farhi2014,zhou2020quantum}, where a classical cost function $C(z)$ is encoded into a corresponding problem Hamiltonian 
\begin{align}
    H_P &= \sum_{z\in \lbrace 0,1 \rbrace^n}C(z)\ketbra{z}{z},
\end{align}
following the energy minimization
\begin{eqnarray}\label{cost-energy}
    E_p(\bm{\beta}, \bm{\gamma}) 
    &= \bra{ \psi_p(\bm{\beta}, \bm{\gamma})} H_{\mathrm{P}} \ket{\psi_p(\bm{\beta}, \bm{\gamma})} \to \min_{\bm{\beta}, \bm{\gamma}}.
\end{eqnarray}
The ansatz \eqref{ion-native-ansatz} avoids one of the critical limitations of standard QAOA circuits: it does not use the problem Hamiltonian exponent $e^{-i\gamma H_P}$, which would typically require a decomposition into a sequence of limited fidelity two-qubit gates. However, this is not free of drawbacks, as the ansatz \eqref{ion-native-ansatz} a priori does not take into account any structure of the problem, which can only be encoded into the choice of hyperparameters $\bm A$. Without this encoding, e.g.~using random generated non-symmetric hyperparameters $\bm A$ produces ansatzes of high expressivity \cite{rabinovich2022ion}. Yet, it can suffer from trainability limitations: even if a solution exists in the variational state space, it might be hard to locate~\cite{holmes2022connecting,AK2022}. 

At the same time, a problem-specific choice of $\bm A$ has been shown to drastically improve the algorithmic performance~\cite{rabinovich2022ion}. Thus, adjusting hyperparameters $\bm{A}$ for each specific problem allows finding a well-trainable ansatz that can find the solution at relatively low circuit depth, which is crucial for near-term quantum devices~\cite{Bharti2022noisy}. In this paper, we focus on how to improve the trainability of ansatz \eqref{ion-native-ansatz} by adjusting its hyperparameters~$\bm{A}$.

\section{\label{sec:heuristics}Heuristic for searching problem-specific hyperparameters}

\vspace{-0.3em}
We seek for trainable ansatz configurations such that transition amplitudes from the initial state to the low energy subspace of a considered problem become large enough starting already from a low circuit depth. We enhance these transition amplitudes by minimizing the energy for a single layer as 
\begin{equation}\label{cost-energy-single-layer}
    E(\bm{\theta},\bm{A}) 
    = \langle \psi_1(\bm{\theta};\bm{A}) | H_{\mathrm{P}} |\psi_1(\bm{\theta}; \bm{A})\rangle \to \min_{\substack{\bm{\theta} \in \Theta \\ \bm{A} \in [-1,1]^{\times n}}}.
\end{equation}
As we demonstrate below, the hyperparameters optimized in this fashion allow to effectively lock the state evolution to a low dimensional subspace. As a result, the ansatz explores a small portion of the Hilbert space that has an overlap with the low energy subspace. This provides a cost landscape with respect to variational parameters favorable for optimization at both depth $p=1$ and larger circuit depths.


\begin{figure*}[t]
    \centering
    \includegraphics[width=1.0\linewidth]{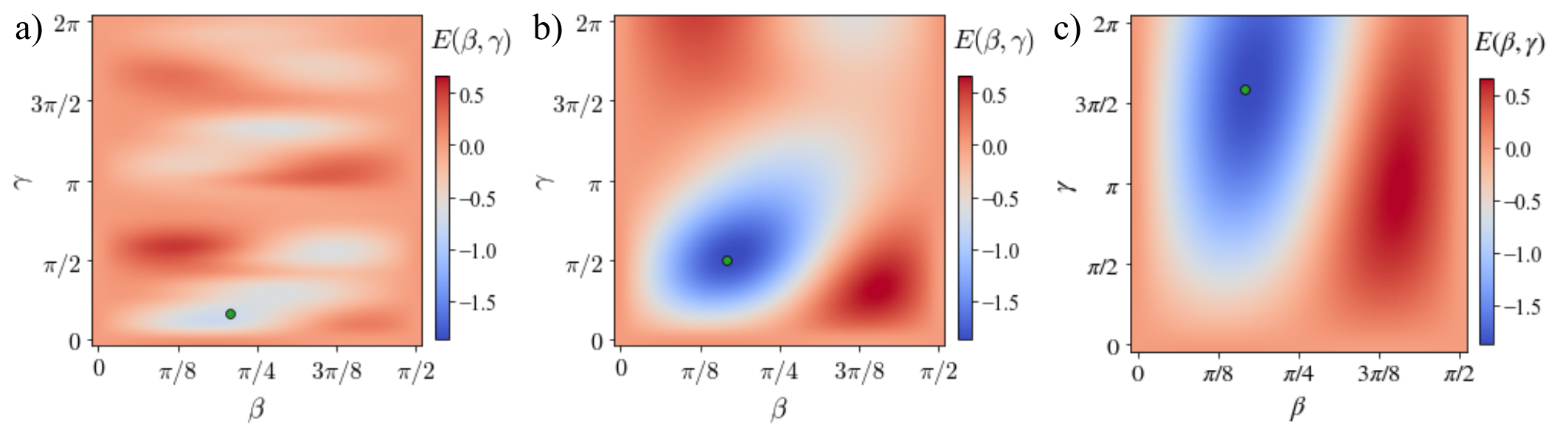}
    \caption{Typical cost landscape for the energy~\eqref{cost-energy-single-layer} of a single-layered ion native ansatz as a function of variational parameters $(\beta,\gamma) \in \Theta$ calculated using different configurations of hyperparameters: a) asymmetric ($A_i = 1$, $\forall i \neq 1$; $A_1 = -0.3$), b) $\bm{A}^*$ trained by our heuristic and c) $\alpha \bm{A}^*$ rescaled by the factor $\alpha = 0.55$ after training. Green points depict the global minima.} 
    \label{fig:scaling}
\end{figure*}

A formal description of the proposed heuristic is given in Algorithm~\ref{alg:heuristics}. The inputs are: the number of qubits $n$, the problem Hamiltonian $H_{\rm P}$, the maximal Rabi frequency $\Omega_{\max}$, the matrix $C_{ij}$ of the phonon contribution to the Ising couplings~\eqref{J-modified}, the maximum allowed block coordinate descent (BCD) iterations $k_{\max}$ and random restarts $m_{\max}$, the tolerance $\varepsilon$ for the convergence criteria, and the tolerance $\delta$ for detecting stagnation in a local minimum. 
The BCD method used for minimizing the energy~\eqref{cost-energy-single-layer} alternately optimizes the cost function with respect to each block of variables $\bm{\theta}$ and $\bm{A}$. One of the main advantages of BCD is that it allows to optimize each block of variables by methods, which take into account specific properties of the problem (see, e.g.,~\cite{Tseng2001}). The algorithm starts by sampling a random initial guess for hyperparameters $\bm{A}_0$ uniformly distributed on $[-1,1]^n$. Then, it proceeds to the main BCD iterations loop, where the energy \eqref{cost-energy-single-layer} is minimized alternately over $\bm{\theta}$ and $\bm{A}$.

On each iteration $k$ of the BCD loop, the current approximation $\bm{A}_k$ for hyperparameters is fixed and  the energy $E(\bm{\theta}, \bm{A}_k)$ is minimized over a pair of variational parameters $\bm{\theta} = (\beta,\gamma) \in \Theta$ (for details, see Appendix~\ref{appendix:numerics}). Once this step is complete, the BCD convergence criteria  
\begin{equation}\label{heuristics-criteria}
    E_k < \varepsilon,
\end{equation}
is checked. Here, $E_k = \min_{\bm{\theta} \in \Theta} E(\bm{\theta}, \bm{A}_k)$ is the optimized energy on the $k$-th iteration. If the criteria \eqref{heuristics-criteria} is met, the algorithm terminates and returns $\bm{A}_k$ as the optimal ansatz configuration~$\bm{A}^*$.  Note that the tolerance $\varepsilon$ in Eq.~\eqref{heuristics-criteria} is chosen empirically. For the combinatorial optimization problems considered later, encoded into traceless Hamiltonians,  $\varepsilon$ was set to $0.5\lambda_0$. 
In these settings the criteria~\eqref{heuristics-criteria} requires a prior knowledge of the ground state energy $\lambda_0$, which is typically not accessible in real experiments. In this case, one should either introduce a new empirical convergence criteria or run the algorithm multiple times to choose the best ansatz configuration.

\begin{algorithm}[H]
\caption{Heuristic for searching a well-trainable hyperparameters configuration of the ion native ansatz~\eqref{ion-native-ansatz} for a specific problem Hamiltonian~$H_{\rm P}$.}
\label{alg:heuristics}
\begin{algorithmic}[1]\onehalfspacing
\Require $n$, $H_{\rm P}$, $\Omega_{\max}$, $C_{ij}$, $k_{\max}$, $m_{\max}$, $\varepsilon$, $\delta$
\Ensure $\bm{A}^*$

\State $L \gets [\,\,]$ \Comment{Empty list to store energies and $\bm{A}$s }

\For {$m$ = 1 to $m_{\max}$}

    \State Sample random guess: $\bm{A}_1 \sim \mathcal{U}(-1,1)^n$
    \State $E_{-1}$ = $\infty$ \Comment{or a large enough number}

    \For {$k$ = 1 to $k_{\max}$}
    
        \State $\bm{\theta}_k \gets \argmin_{\bm{\theta} \in \Theta} E(\bm{\theta},\bm{A}_k)$
        \State $E_k \gets E(\bm{\theta}_k,\bm{A}_k)$

        \If{$E_k < \varepsilon$} \Comment{Check convergence}
            \State \Return $\bm{A}_k$
        \EndIf
        
        \If{$|E_k - E_{-1}| < \delta$} \Comment{Check stagnation}
            \State {\bf break}
        \EndIf

        \State $\bm{A}_{k+1} \gets~\argmin_{\bm{A}\in [-1,1]^n} E(\bm{\theta}_k,\bm{A})$
        \State $E_{-1} \gets E_k$ 
        
    \EndFor
    
    \State append $[E_k, \bm{A}_k]$ to $L$

\EndFor
\State \Return $\bm{A}$ with the smallest energy from $L$

\end{algorithmic}
\end{algorithm}

If the optimization exceeds the maximum allowed number of iterations $k_{\max}$ or gets stuck in a local minimum without significant improvement, it restarts with a new initial guess $\bm{A}_0$ for hyperparameters, while the iteration counter $k$ is reset to zero.
Otherwise, the energy $E(\bm{\theta}_k,\bm{A})$ is minimized over the vector $\bm{A}$ of $n$ hyperparameters, with $\bm{A}_k$ taken as an initial guess. During this process, the block of variational parameters is kept fixed at~$\bm{\theta}=\bm{\theta}_k$. After this step, the next approximation to the optimal configuration of hyperparameters, $\bm{A}_{k+1} = \argmin_{\bm{A}\in [-1,1]^n} E(\bm{\theta}_k,\bm{A})$ is obtained, and the iterations counter $k$ is increased by $1$.

The heuristic continues the BCD iterations, alternately optimizing \eqref{cost-energy-single-layer} with respect to $\bm{\theta}$ and $\bm{A}$ until either the criteria \eqref{heuristics-criteria} is met or the number of random restarts exceeds $m_{\max}$. If the latter occurs, the method returns the hyperparameters that provided the smallest energy~\eqref{cost-energy-single-layer} over all restarts. In practice, only a few iterations were required to reach the desired accuracy in~\eqref{heuristics-criteria} for the problem instances, considered in the next section. 

At Fig.~\ref{fig:scaling}, we examine the effect of ansatz hyperparameters~$\bm{A}^*$ trained by our heuristic on a typical single layer cost landscape $E(\beta,\gamma,\bm{A})$ with respect to variational parameters $(\beta,\gamma) \in \Theta$. At Fig.~\ref{fig:scaling}a we show the landscape~\eqref{cost-energy-single-layer} provided by the asymmetric configuration $\bm{A}$, where $A_i = 1$, $\forall i \neq i'$ and $A_{i'}$ detuned randomly (we set $A_1 = -0.3$) as proposed in \cite{rabinovich2022ion}.
We see that the cost landscape is highly non-convex plagued by multiple local minima. This is a serious limitation for the trainability of variational quantum algorithms~\cite{AK2022}, which only worsens upon increase of the circuit depth ~\cite{RGB2024}.
In contrast, the configuration $\bm{A}^*$ trained by the proposed heuristic provides a more favorable optimization landscape with a single pronounced global minimum (see Fig.~\ref{fig:scaling}b), which is about 2.7 times deeper as compared to the one shown at Fig.~\ref{fig:scaling}a. 

Note that the global minimum near $(\pi/6,\pi/3)$ in Fig.~\ref{fig:scaling}b is located inside a narrow gorge. 
This can limit the ansatz trainability, as the fraction of the parameter space below a certain cost function value becomes exponentially suppressed as $p$ increases~\cite{CSV2021,AHC2022}.
The gorge, however, can be widened by restricting the range of $\gamma$ to the vicinity of the global minimum. If the minimum is located in the region of small~$\gamma$, this can effectively be realized by rescaling $\bm{A}$ by a factor $\alpha \in (0,1)$. Indeed, from the expression for Ising coupling coefficients~\eqref{J-modified}, it is clear that the effective Hamiltonian~\eqref{HI-native} is quadratic with respect to hyperparameters $\bm{A}$. Thus, substituting $A_j \to \alpha A_j$ in~\eqref{J-modified} and taking into account Eq.~\eqref{ion-native-ansatz}, for the single-layered energy~\eqref{cost-energy-single-layer} we obtain
\begin{equation}\label{scaling}
    E(\beta,\gamma, \alpha \bm{A})
    = E(\beta,\alpha^2 \gamma,\bm{A}).
\end{equation}
As a result, the narrow gorge widens as shown at Fig.~\ref{fig:scaling}c, while the energy~\eqref{cost-energy-single-layer} starts to behave almost like a convex function. 

In order to find the optimal scaling factor $\alpha^*$, we construct a cost function $f(\alpha)$ that estimates the fraction of the parameter space $\Theta$ that corresponds to energy~\eqref{cost-energy-single-layer} bellow a certain value as described in Algorithm~\ref{alg:rescaling}. Apart from $n$, $H_{\rm p}$, $\Omega_{\max}$ and $C_{ij}$, the inputs are: the trained hyperparameters $\bm{A}^*$, the energy level $\mu$, and the tolerance $\epsilon$ for keeping the energy minimum within $\Theta$. Additionally, one needs to evaluate the energy landscape $E(\bm{\theta},\bm{A}^*)$ on a $N \times N$ regular grid for $\bm{\theta} \in \Theta$. Its minimum, $E_*^{} = \min_{i,j} E_{ij}$, should also be provided as an input. 
For a given $\alpha$ we evaluate the energy $E(\bm{\theta},\alpha\bm{A}^*)$ for rescaled hyperparameters on the same grid and store it in the form of a matrix~$E_{ij}^\alpha$. Generally, its minimal element, $E_*^\alpha = \min_{i,j} E^\alpha_{ij}$, should be close to~$E_*^{}$. However, it is possible to rescale the energy landscape such that the global minimum moves outside of the parameter space~$\Theta$. 
So, if $|E_*^{} - E_*^\alpha| > \epsilon$, then $f(\alpha)$ returns $-1$. Otherwise, we count the number $M$ of matrix elements $E_{ij}^\alpha$ below an energy level determined by $\mu$, and $f(\alpha)$ returns $M/M_{\rm tot}$, which estimates the fraction of the parameter space $\Theta$. Here, $M_{\rm tot} = N^2$ is the total number of matrix elements. This cost function is maximized to find $\alpha^* = \argmax_{\alpha} f(\alpha)$ (for details, see Appendix~\ref{appendix:numerics}).

\begin{algorithm}[H]
\caption{Cost function $f(\alpha)$ that estimates the fraction of the parameter space $\Theta$ with energies~\eqref{cost-energy-single-layer} below a certain value for a given scaling factor $\alpha$ for the hyperparameters $\bm{A}^*$ trained by the proposed heuristic.}
\label{alg:rescaling}
\begin{algorithmic}[1]\onehalfspacing
\Require $n$, $H_{\rm P}$, $\Omega_{\max}$, $C_{ij}$, $\bm{A}^*$, $\mu$, $\epsilon$, $N$, $E_*$, $\alpha$
\Ensure $f(\alpha)$

\For{$i=1$ to $N$}
\For{$j=1$ to $N$}
\State $E_{ij}^{\alpha} \gets E(\beta_{i}, \gamma_{j}, \alpha\bm{A}^*)$
\EndFor
\EndFor

\State $E_*^{\alpha} \gets \min_{i,j} E_{ij}^{\alpha}$

\If{$|E_*^{} - E_*^{\alpha}| > \epsilon$} \Comment{Check minimum is in $\Theta$}

\State \Return $-1$

\EndIf

\State $M \gets \bigl|\lbrace (i,j) : E_{ij}^\alpha  < \mu E_*^\alpha \rbrace \bigr|$

\State $M_{\rm tot} \gets N^2$

\State \Return $M/M_{\rm tot}$

\end{algorithmic}
\end{algorithm}

\FloatBarrier

Thus, the approach for searching problem-specific hyperparameters of the ion native ansatz \eqref{ion-native-ansatz} consists of two stages. First, the algorithm alternately minimizes the energy~\eqref{cost-energy-single-layer} over the variational and hyper- parameters to find the configuration $\bm{A}^*$ that provides a favorable optimization landscape. Second, the optimal scaling factor $\alpha^*$ for $\bm{A}^*$ is found to stretch the desired part of the cost landscape. 
As a result, the heuristic identifies the configuration $\alpha^* \bm{A}^*$ that excites transitions to the low energy subspace of the considered problem. As we demonstrate in the next section, this improves circuit trainability even for $p > 1$ layers.

Note that the proposed method requires evaluating the energy only for a single layer ($p=1$) ansatz, which is efficient in terms of computational time and is accessible in the experiment. The only potential bottleneck of this heuristic is the optimization of the cost~\eqref{cost-energy-single-layer} over $\bm{A}$ on each BCD iteration. The cost function~\eqref{cost-energy-single-layer} with respect to $\bm{A}$ is $n$-dimensional, while the shape of the optimization landscape is unknown. As $n$ increases, this step of the heuristic can become more challenging for standard optimization techniques.

\section{\label{sec:results}Numerical results}
In this section we demonstrate the performance of heuristically found ansatz configuration at depth $p>1$. We statistically validate its performance across a range of problem instances, demonstrating a reduction of the required circuit depth compared to both problem agnostic hyperparameters and standard QAOA circuit. We provide empirical evidence suggesting that this improvement steams from the expressibility-trainability interplay: the heuristic allows to identify ansatz configurations that lock the evolution to a low dimensional subspace and excite transitions to the low energy states of the problem. Finally, we estimate the number of cost function evaluations for executing the heuristic and show that this number is an order of magnitude smaller than training a variational circuit itself.

\subsection{The problem}
We test the proposed heuristic for the ion native QAOA,~i.e. by solving combinatorial optimization problems with ansatz~\eqref{ion-native-ansatz}. 
As a testbed, we consider random instances of the Sherrington-Kirkpatrick (SK) Hamiltonian~\cite{panchenko2012sherrington},
\begin{equation}\label{H-SK}
    H_{\rm P} = \frac{1}{\sqrt{n}} \sum_{i<j} K_{ij} Z_i Z_j,
\end{equation}
where the all-to-all coupling coefficients $K_{ij}$ are sampled independently from a normal distribution with zero mean and unit variance. A detailed study of the performance of the standard QAOA applied for optimizing the SK model~\eqref{H-SK} can be found in~\cite{Farhi2022}. We have additionally studied the performance of the proposed method for MAX-CUT on random and regular graphs, but no significant difference between the two models has been observed, so the rest of the work focuses on the SK model.

The simulations are performed for system sizes from $n=5$ to $15$ qubits. At each problem size $n$, the matrix $C_{ij}$ of phonon contribution to the Ising couplings~\eqref{J-modified} in the ion Hamiltonian $H_{\rm I}$ is calculated (for details, see Appendix~\ref{appendix:phonons}). For each random instance of \eqref{H-SK}, the proposed heuristic is applied to identify appropriate problem-specific hyperparameters of the ion native ansatz~\eqref{ion-native-ansatz}. The obtained ansatz configuration is then fixed and the energy~\eqref{cost-energy} is minimized at different circuit depths~$p$ using the layerwise training heuristic proposed in \cite{akshay2022circuit}. Note that in order to eliminate degeneracies in the parameter space $\Theta$ due to the $\mathbb{Z}_2$ symmetry, we restrict $\beta_k \in [0,\pi/2)$ in the simulations (see, e.g., \cite{wauters2020polynomial}).

The performance of the algorithm is evaluated using two different metrics. The first metric is the normalized approximation ratio (see, e.g., \cite{pagano2020quantum}),
\begin{equation}\label{approx-ratio}
    r = \frac{\lambda_{\max} - E_p(\bm{\beta}^*, \bm{\gamma}^*) }{\lambda_{\max} - \lambda_0},
\end{equation}
where $\lambda_{\max}$ is the energy of the highest excited state of $H_{\rm P}$, and $(\bm{\beta}^*, \bm{\gamma}^*)$ are the optimal variational parameters obtained by optimizing the energy \eqref{cost-energy}. 
The second one is the overlap $g(\psi)$ of a state $\ket{\psi} \equiv \ket{\psi_p(\bm{\beta}^*, \bm{\gamma}^*)}$ with the $d$-degenerate ground space of $H_{\rm P}$ (see, e.g.,~\cite{akshay2022circuit}),
\begin{equation}\label{overlap-1}
    g(\psi) = \sum_{j=1}^d |\bra{\phi_j}\ket{\psi}|^2,
\end{equation}
where $\ket{\phi_j}$ are the corresponding eigen basis states. We consider a simulation run to be successful, if the resulting overlap \eqref{overlap-1} surpasses the threshold $g(\psi) > 0.5$. Note that a high approximation ratio does not necessarily guarantee a certain overlap with the ground state. According to the stability lemma \cite{biamonte2021universal}, the relation between the two strongly depends on the spectral gap $\Delta$, which can be small for specific problem Hamiltonians $H_{\rm P}$. 

\subsection{Effect of hyperparameters on ansatz trainability}\label{subsect:trainability}
\begin{figure}[ht!]
    \centering
    \includegraphics[width=0.9\linewidth]{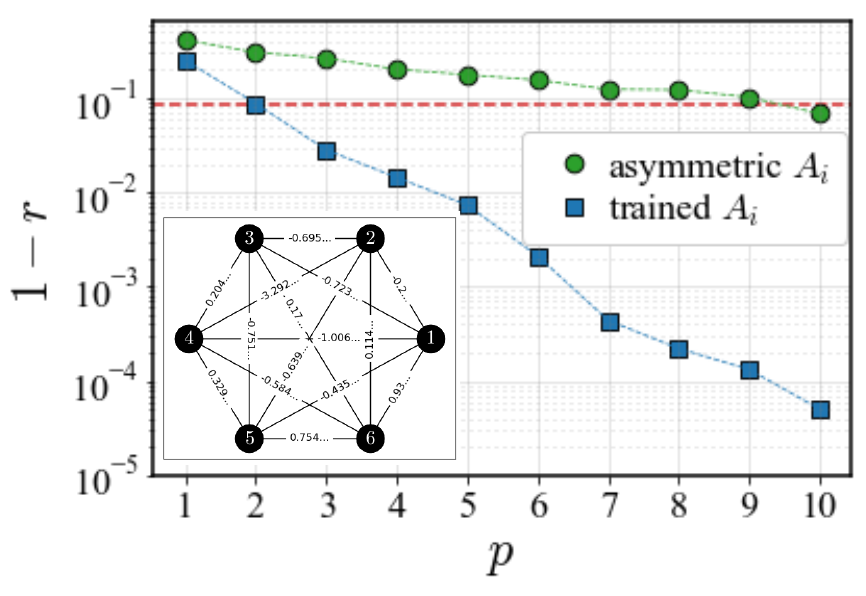}
    \caption{Performance of the ion native QAOA in terms of the fractional error $1-r$ as a function of the circuit depth~$p$. The results are obtained for (i) the asymmetric configuration ($A_i = 1$, $\forall i \neq 1$; $A_1 = -0.3$), shown in green, and (ii) found by the heuristic, shown in blue. Red dashed line depicts the threshold, which guarantees the ground state overlap $g(\psi) = 0.5$ from the stability lemma \cite{biamonte2021universal}. The inset depicts the coefficients $K_{ij}$ for the considered SK problem instance~\eqref{H-SK}.} 
    \label{fig:single_instance}
\end{figure}

We begin by demonstrating the role of an appropriate problem-specific choice of hyperparameters $\bm{A}$ on the trainability of the ion native circuit. 
For a randomly generated SK instance (see the inset at Fig.~\ref{fig:single_instance}), we compare the QAOA performance for two different configurations of $\bm{A}$. The first is the asymmetric one, where $A_i = 1$, $\forall i \neq i'$ and $A_{i'}$ detuned randomly (we set $A_1 = -0.3$). This configuration is known to be able to minimize all instances of the simplified model~\eqref{H-SK} with $K_{ij} = \pm 1$ for $n=6$ qubits using the QAOA circuit depth of up to $p=20$ layers \cite{rabinovich2022ion}. The second one is a problem-specific configuration obtained from the proposed heuristic. The respective cost landscapes \eqref{cost-energy-single-layer} for a single-layered ansatz are presented at Figs.~\ref{fig:scaling}a and~\ref{fig:scaling}c.

Fig.~\ref{fig:single_instance} shows the optimization results for the considered hyperparameter configurations at different circuit depths. 
Comparing the fractional error $1-r$ in these configurations, one can observe a dramatic difference in the algorithmic performance. 
For the asymmetric configuration of $\bm{A}$, the fractional error $1-r$ slowly decreases with $p$ reaching the threshold $g(\psi) > 0.5$ at $p=10$ layers. In comparison, the problem-specific configuration of $\bm{A}$ obtained by the proposed heuristic drastically improves the performance of ion native QAOA. Here, the fractional error $1-r$ rapidly decreases with $p$ to about $10^{-5}$ at $p=10$. Moreover, in this case already $p = 2$ layers of the QAOA circuit are sufficient to reach the threshold $g(\psi)>0.5$. Evidently, using appropriate problem-specific hyperparameters, one can drastically reduce the circuit depth required to reach a desired accuracy.

\subsection{Statistical evaluation on random instances}
\label{subsect:stats}

In this section we systematically investigate the performance of the proposed heuristic on random instances of the SK model~\eqref{H-SK} in the range from 5 to 10 qubits. 
As we show in Appendix~\ref{appendix:stateprep}, for all problems of a given size, the required ground state overlap $g(\psi) > 0.5$ can be achieved using the same circuit depth. This fact is demonstrated by solving the state preparation problem, where $p=2$ layers of the ion native ansatz~\eqref{ion-native-ansatz} are sufficient to reach the desired overlap threshold for $n=5$--$10$. 
This, however, clearly ignores how hard it is to train such circuits and adjust hyperparameters when considering the Hamiltonian minimization. Thus, the solution at depth $p=2$, while exists, might be practically unattainable, therefore we relax it by considering deeper circuits when minimizing SK instances.

For each number of qubits, 100 random instances of the SK model are sampled with $K_{ij} \sim \mathcal{N}(0,1)$. 
Several training cycles are performed for the sampled instances. Each cycle consists of identifying hyperparameters for every specific instance, 
followed by the Hamiltonian minimization using the obtained ansatzes. 
After each cycle, we evaluate the fraction of solved instances using a $p=n$ layer circuit. 
Each subsequent cycle is run only for the remaining unsolved instances.

\begin{figure}[ht!]
    \centering
    \includegraphics[width=0.9\linewidth]{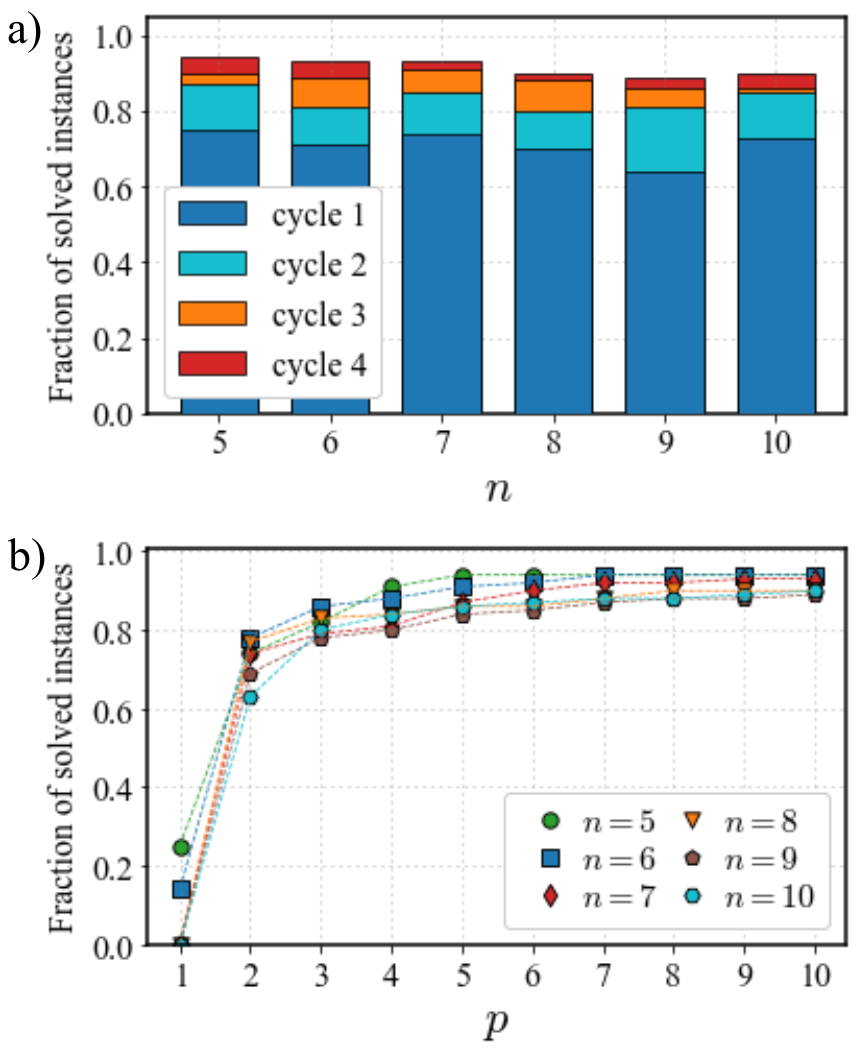}
    \caption{Fraction of SK instances~\eqref{H-SK} solved with the ion native QAOA for different system sizes $n$ and plotted a) after each cycle of training $p=n$ layered QAOA circuit and b) after the 4-th cycle of training as a function of the QAOA circuit depth $p$. } 
    \label{fig:frac}
\end{figure}

Fig.~\ref{fig:frac}a demonstrates the fraction of solved SK instances evaluated after each cycle of training for different number of qubits $n$. As one can see, after a single training cycle, $64-75\%$ of sampled instances are solved. This is improved by running additional cycles: after running four of them, the fraction of solved instances increases to $89-94\%$. 
Note that reaching the same fraction of solved instances might require a number of training cycles that increases with the system size $n$. 
This can be explained by the fact that it becomes harder to optimize the energy~\eqref{cost-energy-single-layer} for larger $n$.

Fig.~\ref{fig:frac}b presents the numerical results for the fraction of solved SK instances calculated as a function of the QAOA circuit depth for $n=5$--10 qubits and $p \leq 10$ layers. These results are obtained for hyperparameters $\bm{A}$ computed in four cycles of training for random instances as described above. 
For all considered problem sizes $n$, the fraction of solved instances experiences a sharp jump at $p=2$ --- in accordance with the results of state preparation --- increasing up to 89--94\% for circuit depth $p=10$. Interestingly, for every problem size $n$ it can be observed that $p=4$ layers of the ion native ansatz are sufficient to solve most instances (greater than $80\%$). We have also checked the heuristic scalability by running the same set of experiments for $n=15$ qubits, where $50\%$ of the instances have been solved after a single cycle of training, increasing to $76\%$ after four cycles. This, together with the results illustrated in Fig.~\ref{fig:frac}a, demonstrates no significant degradation of the heuristic performance upon the system size increase.

\begin{figure}[ht!]
    \centering
    \includegraphics[width=0.9\linewidth]{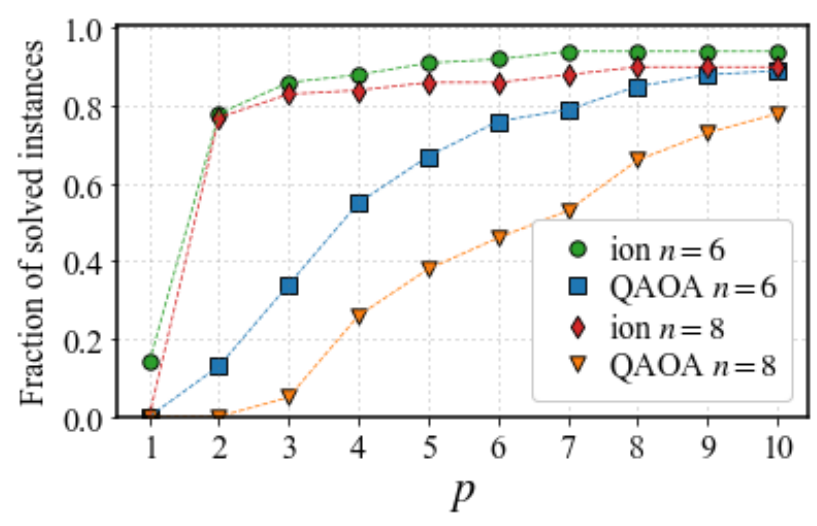}
    \caption{Fraction of SK instances \eqref{H-SK} for $n=6$ and 8 qubits solved by the ion native and standard QAOA as a function of circuit depth $p$. The results for the ion-based QAOA are obtained using problem-specific hyperparameters found by the proposed heuristic.} 
    \label{fig:qaoa}
\end{figure}

Finally, Fig.~\ref{fig:qaoa} compares the performance of the ion native QAOA with the best found hyperparameters and standard QAOA, solving the same pool of sampled SK instances with the same settings of the layerwise optimization heuristic (Appendix~\ref{appendix:numerics}).
Two key observations can be made. First, for a given problem size, the ion native QAOA strongly outperforms the standard one in terms of the fraction of solved instances. Compared to the ion native implementation, standard QAOA exhibits a much slower performance improvement as depth $p$ increases. 
Second, the ion native QAOA experiences a lesser performance degradation as the system size increases. 

\FloatBarrier
\vspace{-1em}
\subsection{Trainability and expressibility}
The expressibility of a parameterized quantum circuit is defined by how uniformly it explores the unitary space~\cite{SJA2019}. Recent studies demonstrate that the expressibility of an ansatz can be negatively correlated to the magnitude of cost function gradients, thus limiting the circuit trainability~\cite{holmes2022connecting}. 
Here, we provide empirical arguments that the ion native ansatz~\eqref{ion-native-ansatz} with heuristically found hyperparameters possesses limited expressibility compared to the other ansatze considered: instead, it locks the evolution into a subspace that overlaps with low energy states.

To evaluate the expressibility we use a descriptor based on the Kullback-Leibler (KL) divergence~\cite{SJA2019},
\begin{equation}\label{DKL}
    D_{\rm KL} = \int_0^1 \rho_F(x) \,\log {\frac {\rho_F(x)}{\rho_{\mathrm{Haar}}(x)}}\,{\mathrm{d}} x,
\end{equation}
between the probability density function $\rho_F(x)$ for the distribution of fidelities $F = \bigl| \bra{\psi(\bm{\theta}_1)}\ket{\psi(\bm{\theta}_2)} \bigr|^2$ sampled uniformly from the ansatz, and for Haar random states
\begin{equation}\label{Haar}
    \rho_{\mathrm{Haar}}(x) = (N-1)(1 - x)^{N-2}.
\end{equation}
In Eq.~\eqref{Haar}, $N$ is the dimension of the Hilbert space of quantum states. The ansatz with smaller value of $D_{\rm KL}$ has a higher
expressibility. 
The KL divergence~\eqref{DKL} is evaluated in a range of  ansatz circuit depths~$p$. At each depth~$p$, $10^5$ fidelities are produced from the ion native ansatz~\eqref{ion-native-ansatz} by sampling variational parameters $(\bm{\beta}, \bm{\gamma})$ uniformly from $\Theta$. For the Haar density \eqref{Haar}, $N=2^{n-1}$ is used to account for the $\mathbb{Z}_2$ symmetry of the QAOA prepared states, $X^{\otimes n}\ket{\psi} = \ket{\psi}$. 

\begin{figure}[b]
    \centering
    \includegraphics[width=0.88\linewidth]{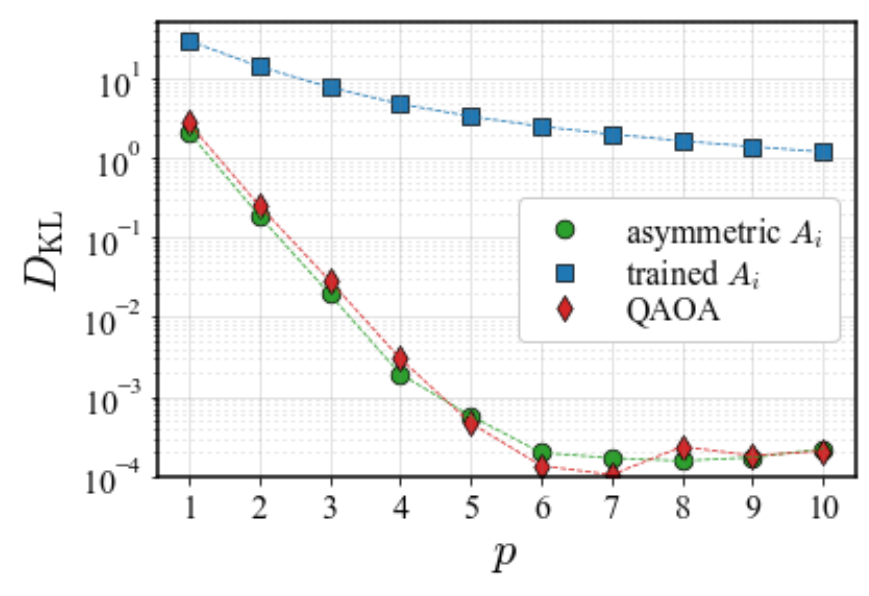}
    \caption{KL divergence~\eqref{DKL} as a function of the circuit depth $p$ for $n=6$ qubits. The results are obtained for the ion native ansatz~\eqref{ion-native-ansatz} using the asymmetric configuration $\bm{A}$ (green), using $\bm{A}^*$ found by the heuristic (blue), and for standard QAOA (red). The considered SK instance is shown in the inset of Fig.~\ref{fig:single_instance}a. Note that lower values of $D_{\rm KL}$ correspond to a more expressible circuit.} 
    \label{fig:expressibility}
\end{figure}

Fig.~\ref{fig:expressibility} illustrates expressibility of the ion native ansatz~\eqref{ion-native-ansatz} for $n=6$ qubits for the asymmetric and trained configurations of $\bm{A}$ from Sec.~\ref{subsect:trainability}. Note that the descriptor~\eqref{DKL} does not depend on a specific problem instance. We observe that the KL divergence for the asymmetric configuration rapidly decreases with $p$ to about $D_{\rm KL} \sim 10^{-4}$ and saturates starting from $p=6$ layers.
For the trained hyperparameters, on the other hand, the ansatz expressibility is strongly limited: the KL divergence takes typical values $D_{\rm KL} \gtrsim 1$ and slowly decreases with $p$. In other words, by training the hyperparameters we identify a problem-specific ansatz, designed to prepare a fraction of the Hilbert space, containing the solution. 
This naturally limits the expressibility of the designed ansatz, locking it to the space in the vicinity of the solution, but allows to enhance its trainability.

To further support this argument, we investigate the subspace spanned by the ansatz states $\ket{\psi_p(\bm{\theta})}$ obtained for random variational parameters $\bm{\theta}$ sampled uniformly from $\Theta$. We quantify their linear independence by computing the singular values $\sigma_k$ of a $(K \times 2^n)$-dimensional matrix constructed from $K$ of these random quantum states. In the case of completely independent (i.e.~orthogonal vectors), the distribution of singular values would be flat. On the contrary, if the vectors tend to cluster within some subset of the space, the distribution is expected to exhibit a sharp drop, as the vectors would be almost linearly dependent.

\begin{figure}[ht!]
    \centering
    \includegraphics[width=0.98\linewidth]{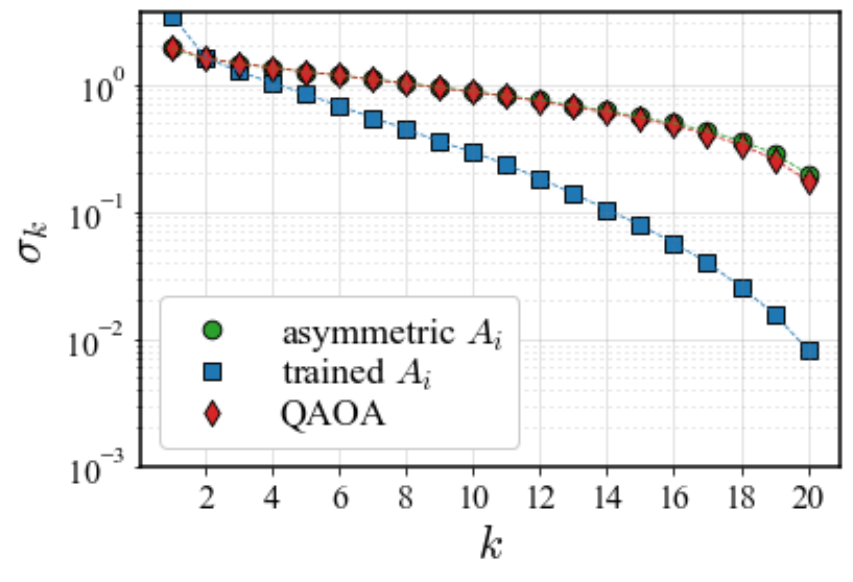}
    \caption{Singular values that quantify the linear independence of quantum states prepared by the ion native ansatz~\eqref{ion-native-ansatz} using the asymmetric configuration of $\bm A$ (green), heuristically found $\bm A^*$ (blue), and for standard QAOA (red). The results are obtained using the circuit depth of $p=10$ layers for $n=6$ qubits. The values $\sigma_k$ are calculated from the singular value decomposition of a matrix composed of $K=20$ ansatz states $\ket{\psi_p(\bm{\theta})}$ for random variational parameters $\bm{\theta}$ sampled uniformly from $\Theta$. The blue and red curves are averaged over SK problem instances used in Sec.~\ref{subsect:stats}.}
    \label{fig:singulars}
\end{figure}

Fig.~\ref{fig:singulars} shows the singular values $\sigma_k$ calculated for the ion native QAOA circuit~\eqref{ion-native-ansatz} of depth $p=10$ as well as for standard QAOA for $n=6$ qubits. The curves of singular values $\sigma_k$ for standard QAOA and its ion native counterpart with trained hyperparameters are averaged over the SK instances sampled in Sec.~\ref{subsect:stats}. As expected, the curves for standard QAOA and ion native ansatz with asymmetric hyperparameters $\bm A$ exhibit a relatively slow decay of singular values, in agreement with their close proximity to the Haar distribution (see Fig.~\ref{fig:expressibility}). However, the ansatz with heuristically trained hyperparameters exhibits a much faster decay of $\sigma_k$, which implies the concentration of the prepared state vectors in a narrow subset of the Hilbert space. By truncating these singular values at a certain threshold, one can argue that the entire evolution is largely contained in a subspace of the entire space. In other words, we observe that the ansatz is mostly locked to a subspace that, as evidenced by the previous subsection, overlaps with the low energy subspaces of the considered problem. This overlap is a result of Algorithm 1 of the heuristic, which ensures that the ansatz~\eqref{ion-native-ansatz} can excite transitions to the low energy subspace. The Algorithm 2 then (if necessary) scales  magnitudes of $\bm A^*$ down, effectively reducing the accessible phases of $\bm \gamma$, as per Eq.~\eqref{scaling}. This reduces the expressibility and locks the evolution, while keeping the low energy subspace accessible to the ansatz.

\FloatBarrier

\FloatBarrier
\vspace{-1em}

\subsection{Computational resources}

For noisy intermediate scale quantum (NISQ) devices, a number of measurements (or shots) required for estimating the mean value of an observable can be a serious limitation for the overall algorithm runtime~\cite{Bharti2022noisy}. In other words, a large number of the cost function evaluations can be very expensive. Thus, we estimate this number for the simulations in Sec.~\ref{subsect:stats} and investigate the scaling of the computational resources required by the proposed heuristic with respect to system size $n$. 

Table~\ref{tab:nfev} shows the number of cost function evaluations $N_{\rm fev}$ used to train and rescale hyperparameters $\bm{A}$ of the ion native ansatz~\eqref{ion-native-ansatz} using the heuristic for different system sizes~$n$.
The results are compared to the number $N_{\rm fev}$ of function evaluations used by the layerwise heuristic optimization routine to minimize the energy~\eqref{cost-energy} for the ion native QAOA circuit. All numbers are averaged over the same pool of random SK instances sampled for each $n$ as described in Sec.~\ref{subsect:stats}.

For training hyperparameters in the BCD loop, we observe that $N_{\rm fev}$ increases with the system size $n$. This is in agreement with the growth of the dimension of the optimization problem~\eqref{cost-energy-single-layer} upon the increase in the number of hyperparameters $\bm{A}$. However, this training process involves only a single layer circuit, whose execution is relatively cheap in terms of computational resources. 

For rescaling hyperparameters, the number of cost function evaluations does not scale with $n$ and remains at about $N_{\rm fev} \sim 5000$. The rescaling procedure consists of maximizing the desired part of the cost landscape represented as a single-variable function $f(\alpha)$ as explained in Sec.~\ref{sec:heuristics}. Every evaluation of $f(\alpha)$ involves a scan of the cost landscape for single-layer energy~\eqref{cost-energy-single-layer} over a $20 \times 20$ discretization grid with respect to a pair of variational parameters $\bm{\theta} = (\beta,\gamma)$ (for details, see Appendix~\ref{appendix:numerics}). Our numerics suggest that the general shape of this landscape does not depend on $n$. Note that the number of cost evaluations for rescaling hyperparameters can be further reduced by implementing a more economical method to evaluate the energy landscape, for example, based on coarse grid interpolation techniques. 

For comparison, we show the number of cost evaluations $N_{\rm fev}$ used by the layerwise training to optimize the ion native QAOA circuit~\eqref{ion-native-ansatz} at depth $p=n$. The circuit optimization was performed using the best found hyperparameters for random SK instances. As one can see, for optimizing a $p=n$ layered QAOA circuit, one needs to perform an order of magnitude more cost function evaluations compared to training ansatz hyperparameters for any specific system size $n$. Moreover, these cost function evaluations are also more expensive since they involve the execution of deeper circuits with multiple layers. This makes our heuristic even more attractive in potential applications, as the small price for training hyperparameters pays off by a significant reduction of the required QAOA circuit depth as shown in Sec.~\ref{subsect:trainability}. As a result, the heuristic has great potential for reducing the overall algorithm runtime on ion hardware.

\begin{table}[ht!]
\centering
\caption{Number of cost function evaluations $N_{\rm fev}$ \eqref{cost-energy-single-layer} used by the proposed heuristic to train and rescale hyperparameters of the ion native ansatz for random SK instances for different system sizes $n$. The results are compared to the number $N_{\rm fev}$ of function evaluations used by the layerwise heuristic \cite{akshay2022circuit} optimization routine to minimize the energy \eqref{cost-energy} for the $p=n$ layered ion-based QAOA circuit with the best found hyperparameters. All numbers are averaged over 100 random SK instances for each $n$.}
\label{tab:nfev}
\begin{tabular}{|c|c|c|c|}
\hline
$n$ & \begin{tabular}[c]{@{}c@{}}Training\\ $\bm{A}$\end{tabular} & \begin{tabular}[c]{@{}c@{}}Rescaling\\ $\bm{A}$\end{tabular} & \begin{tabular}[c]{@{}c@{}}Training QAOA\\ of $p=n$ layers\end{tabular} \\ \hline
5    & 1771.67 & 5140.0  & 29319.05   \\ \hline
6    & 2111.70 & 4592.0  & 44090.79  \\ \hline
7    & 4070.95 & 5120.0  & 73373.49  \\ \hline
8    & 4284.23 & 4956.0  & 113658.42  \\ \hline
9    & 6333.36 & 4540.0  & 156995.73  \\ \hline
10   & 9551.13 & 4544.0 & 217381.61   \\ \hline
\end{tabular}
\end{table}


\FloatBarrier

\vspace{-0.6em}

\section{\label{sec:conclusions}Conclusions}
Digital-analog quantum circuits provide a promising tool for reducing algorithms dependence on limited fidelity entangling gates. A notable example of this approach is the ion based quantum architecture, as it naturally supports the long range all-to-all interaction \cite{monroe2021programmable}. The latter can be controlled on a real quantum device by varying intensities of the applied laser fields. As evidenced by a recent proposal of the ion native ansatz for combinatorial  optimization~\cite{rabinovich2022ion}, the algorithmic performance strongly depends on the choice of interaction parameters (called ansatz hyperparameters $\bm A$). Identifying a suitable configuration of these hyperparameters, however, can itself be a challenging problem. 

In this work, we addressed the problem of identifying hyperparameters of the ion native ansatz suitable for minimizing specific problem Hamiltonians. We proposed a heuristic for searching a well-trainable ansatz configuration based on a single layer optimization. The heuristic uses the block coordinate descent method for minimizing the energy alternately over variational and hyper- parameters. The obtained hyperparameters are then rescaled by a factor which maximizes the fraction of variational parameters space below a certain energy threshold. This allows to eliminate a potential narrow gorge problem in the optimization landscape~\cite{AHC2022}. The proposed heuristic has a significant advantage as it requires evaluating the energy only for a single layer, which has a low computational cost and is relatively easy to access in an experiment. 

We systematically studied the heuristic performance on random instances of the Sherrington-Kirkpatrick (SK) Hamiltonian. We demonstrated that our heuristic obtains ansatz hyperparameters that provide a favorable single layer cost landscape, where the energy behaves almost like a convex function with a single global minimum located in a pronounced well. 
For system sizes up to $n=15$, we generated 100 random SK instances and studied the fraction of the instances that was solved using the ion-based QAOA with problem-specific hyperparameters. We showed that this fraction, evaluated at the circuit depth $p=n$, increases after each training cycle. After four of these cycles, $89$-$94\%$ of problem instances were solved at $p=n$ layers for $n$ from 5 to 10. Moreover, we calculated the fraction of solved SK instances as a function of $p$ and found that more than 80\% of all instances can be solved using no more than $p=4$ layers for the considered system sizes. Identical experiments for $n=15$ qubits demonstrated no significant performance degradation. Similar results were also observed for MAX-CUT on random and regular graphs.

Finally, the obtained numerical results were compared to the performance of standard QAOA. For a given problem size, we found that the ion native QAOA strongly outperforms the standard one in terms of the fraction of solved instances. Compared to the standard implementation, ion native QAOA, enhanced with the proposed heuristic, exhibits a much faster performance improvement with circuit depth. Moreover, the ion native QAOA experiences a lesser performance degradation as the system size increases.


The mechanism behind the proposed heuristic can be understood as follows. By minimizing the expectation value at depth one, it identifies an ansatz capable of exciting transitions to the low energy states. For deeper circuits, this, as empirically observed, effectively locks the evolution to a subspace that overlaps with the low energy states. As a result, the prepared state can be close to the ground state --- allowing us to classify an instance as solved --- or close to one of the other low energy states, which would be considered a suboptimal solution. Indeed, we observed that training cycles which did not succeed in preparing the true solution were approximating one of the low energy excited states.
In this work, we focused on the preparation of the ground state, yet in the regime of a large scale combinatorial optimization these suboptimal solutions can be considered acceptable. Thus, the heuristic can be expected to be useful for large scale problem instances as well. 
This also allows us to conjecture that a similar approach can be used for other problems, not limited to combinatorial optimization.

We estimated the computational resources required to run the proposed heuristic. We showed that the number of cost function evaluations used by the heuristic to train and rescale hyperparameters of the ion native ansatz is an order of magnitude lower as compared to optimizing variational parameters of a QAOA circuit itself. Once again, the heuristic involves only executions of a single layered ansatz, which is relatively cheap in computational time. 
As a result, the small price for training
hyperparameters pays off by a significant reduction of the
required QAOA circuit depth. This allows to potentially reduce the
overall algorithm runtime on a real ion hardware.
We believe these results demonstrate the advantages of the ion native ansatz which, in addition to being naturally implementable on a trapped ions quantum computer, can be tuned to drastically improve the algorithmic performance. This is crucial for implementing variational quantum algorithms on near-term quantum computers. 

\vspace{1em}
\section*{acknowledgments}
We are grateful to Evgeny Anikin for providing the code for simulating a chain of trapped ions \cite{Anikin2025fast}. We acknowledge Andrey Kardashin for sharing tools useful for our numerical simulations. We also acknowledge the usage of the computational resources at the Skoltech supercomputer ``Zhores''~\cite{Zhores}. 
The work was supported by Rosatom in the framework of the Roadmap for Quantum computing (Contract No. 868-1.3-15/15-2021 dated October 5).

\section*{Data Availability Statement}

The data and code that support the findings of
this study are publicly available~\cite{github} (\url{https://github.com/pargv/heuristic_trapped_ions}).
\FloatBarrier

\bibliography{main.bib}

@article{kokail2019self,
  title={Self-verifying variational quantum simulation of lattice models},
  author={Kokail, Christian and Maier, Christine and van Bijnen, Rick and Brydges, Tiff and Joshi, Manoj K and Jurcevic, Petar and Muschik, Christine A and Silvi, Pietro and Blatt, Rainer and Roos, Christian F and others},
  journal={Nature},
  volume={569},
  number={7756},
  pages={355--360},
  year={2019},
  publisher={Nature Publishing Group UK London},
  doi = {https://doi.org/10.1038/s41586-019-1177-4}
}

@article{akshay2020reachability,
  title={Reachability deficits in quantum approximate optimization},
  author={Akshay, Vishwanathan and Philathong, Hariphan and Morales, Mauro ES and Biamonte, Jacob D},
  journal={Phys. Rev. lett.},
  volume={124},
  number={9},
  pages={090504},
  year={2020},
  publisher = {American Physical Society},
  doi = {10.1103/PhysRevLett.124.090504},
  url = {https://link.aps.org/doi/10.1103/PhysRevLett.124.090504}
}

@article{claes2021instance,
  doi = {10.22331/q-2021-09-15-542},
  url = {https://doi.org/10.22331/q-2021-09-15-542},
  title = {Instance {I}ndependence of {S}ingle {L}ayer {Q}uantum {A}pproximate {O}ptimization {A}lgorithm on {M}ixed-{S}pin {M}odels at {I}nfinite {S}ize},
  author = {Claes, Jahan and {van Dam}, Wim},
  journal = {{Quantum}},
  issn = {2521-327X},
  publisher = {{Verein zur F{\"{o}}rderung des Open Access Publizierens in den Quantenwissenschaften}},
  volume = {5},
  pages = {542},
  month = sep,
  year = {2021}
}

@article{wauters2020polynomial,
  title = {Polynomial scaling of the quantum approximate optimization algorithm for ground-state preparation of the fully connected $p$-spin ferromagnet in a transverse field},
  author = {Wauters, Matteo M. and Mbeng, Glen B. and Santoro, Giuseppe E.},
  journal = {Phys. Rev. A},
  volume = {102},
  issue = {6},
  pages = {062404},
  numpages = {10},
  year = {2020},
  month = {Dec},
  publisher = {American Physical Society},
  doi = {10.1103/PhysRevA.102.062404},
  url = {https://link.aps.org/doi/10.1103/PhysRevA.102.062404}
}

@article{zhou2020quantum,
  title = {{Quantum Approximate Optimization Algorithm: Performance, Mechanism, and Implementation on Near-Term Devices}},
  author = {Zhou, Leo and Wang, Sheng-Tao and Choi, Soonwon and Pichler, Hannes and Lukin, Mikhail D.},
  journal = {Phys. Rev. X},
  volume = {10},
  issue = {2},
  pages = {021067},
  numpages = {23},
  year = {2020},
  month = {Jun},
  publisher = {American Physical Society},
  doi = {10.1103/PhysRevX.10.021067},
  url = {https://link.aps.org/doi/10.1103/PhysRevX.10.021067}
}

@ARTICLE{morales2020universality,
       author = {{Morales}, M~E~S and {Biamonte}, J~D and {Zimbor{\'a}s}, Z},
        title = {On the universality of the quantum approximate optimization algorithm},
      journal = {Quantum Information Processing},
         year = 2020,
       volume = {19},
       number = {9},
          eid = {291},
        pages = {291},
          doi = {10.1007/s11128-020-02748-9}
}

@article{akshay2021parameter,
  title={Parameter concentrations in quantum approximate optimization},
  author={Akshay, Vishwanathan and Rabinovich, Daniil and Campos, Ernesto and Biamonte, Jacob},
  journal={Phys. Rev. A},
  volume={104},
  number={1},
  pages={L010401},
  year={2021},
  publisher = {American Physical Society},
  doi = {10.1103/PhysRevA.104.L010401},
  url = {https://link.aps.org/doi/10.1103/PhysRevA.104.L010401}
}

@article{campos2021training,
  title={Training saturation in layerwise quantum approximate optimization},
  author={Campos, Ernesto and Rabinovich, Daniil and Akshay, Vishwanathan and Biamonte, J},
  journal={Phys. Rev. A},
  volume={104},
  number={3},
  pages={L030401},
  year={2021},
  publisher = {American Physical Society},
  doi = {10.1103/PhysRevA.104.L030401},
  url = {https://link.aps.org/doi/10.1103/PhysRevA.104.L030401}
}

@misc{farhi2014quantumBOCP,
      title={{A Quantum Approximate Optimization Algorithm Applied to a Bounded Occurrence Constraint Problem}}, 
      author={Farhi, Edward and Goldstone, Jeffrey and Gutmann, Sam},
      year={2014},
      eprint={1412.6062},
      archivePrefix={arXiv},
      primaryClass={quant-ph},
      url={https://arxiv.org/abs/1412.6062}, 
}

@misc{lloyd2018quantum,
      title={{Quantum approximate optimization is computationally universal}}, 
      author={Lloyd, Seth},
      year={2018},
      eprint={1812.11075},
      archivePrefix={arXiv},
      primaryClass={quant-ph},
      url={https://arxiv.org/abs/1812.11075}, 
}

@article{wang2018quantum,
  title={Quantum approximate optimization algorithm for maxcut: A fermionic view},
  author={Wang, Zhihui and Hadfield, Stuart and Jiang, Zhang and Rieffel, Eleanor G},
  journal={Phys. Rev. A},
  volume={97},
  number={2},
  pages={022304},
  year={2018},
  publisher = {American Physical Society},
  doi = {10.1103/PhysRevA.97.022304},
  url = {https://link.aps.org/doi/10.1103/PhysRevA.97.022304}
}

@article{rabinovich2022progress,
  title={Progress towards analytically optimal angles in quantum approximate optimisation},
  author={Rabinovich, Daniil and Sengupta, Richik and Campos, Ernesto and Akshay, Vishwanathan and Biamonte, Jacob},
  journal={Mathematics},
  volume={10},
  number={15},
  pages={2601},
  year={2022},
  publisher={MDPI},
  URL = {https://www.mdpi.com/2227-7390/10/15/2601},
  doi = {10.3390/math10152601},
}

@article{akshay2022circuit,
  title={Circuit depth scaling for quantum approximate optimization},
  author={Akshay, Vishwanathan and Philathong, H and Campos, E and Rabinovich, Daniil and Zacharov, Igor and Zhang, Xiao-Ming and Biamonte, Jacob D},
  journal={Phys. Rev. A},
  volume={106},
  number={4},
  pages={042438},
  year={2022},
  publisher = {American Physical Society},
  doi = {10.1103/PhysRevA.106.042438},
  url = {https://link.aps.org/doi/10.1103/PhysRevA.106.042438}
}

@article{campos2024depth,
  title={Depth scaling of unstructured search via quantum approximate optimization},
  author={Campos, Ernesto and Rabinovich, Daniil and Uvarov, Alexey},
  journal={Phys. Rev. A},
  volume={110},
  number={1},
  pages={012428},
  year={2024},
  publisher = {American Physical Society},
  doi = {10.1103/PhysRevA.110.012428},
  url = {https://link.aps.org/doi/10.1103/PhysRevA.110.012428}
}

@article{zhuang2024hardware,
  title={Hardware-efficient variational quantum algorithm in a trapped-ion quantum computer},
  author={Zhuang, J-Z and Wu, Y-K and Duan, L-M},
  journal={Phys. Rev. A},
  volume={110},
  number={6},
  pages={062414},
  year={2024},
  publisher = {American Physical Society},
  doi = {10.1103/PhysRevA.110.062414},
  url = {https://link.aps.org/doi/10.1103/PhysRevA.110.062414}
}

@Article{richerme2014non,
  author    = {Philip Richerme and Zhe-Xuan Gong and Aaron Lee and Crystal Senko and Jacob Smith and Michael Foss-Feig and Spyridon Michalakis and Alexey V. Gorshkov and Christopher Monroe},
  journal   = {Nature},
  title     = {Non-local propagation of correlations in quantum systems with long-range interactions},
  year      = {2014},
  month     = {jul},
  number    = {7508},
  pages     = {198--201},
  volume    = {511},
  doi       = {10.1038/nature13450},
  publisher = {Springer Science and Business Media {LLC}},
}

@Article{senko2014coherent,
  author    = {Senko, C. and Smith, J. and Richerme, P. and Lee, A. and Campbell, W. C. and Monroe, C.},
  title     = {Coherent imaging spectroscopy of a quantum many-body spin system},
  year      = {2014},
  month     = {jul},
  number    = {6195},
  pages     = {430--433},
  volume    = {345},
  doi       = {10.1126/science.1251422},
  publisher = {American Association for the Advancement of Science ({AAAS})},
  journal = {Science},
}

@article{gentini2020noise,
  title={Noise-resilient variational hybrid quantum-classical optimization},
  author={Gentini, Laura and Cuccoli, Alessandro and Pirandola, Stefano and Verrucchi, Paola and Banchi, Leonardo},
  journal={Phys. Rev. A},
  volume={102},
  number={5},
  pages={052414},
  year={2020},
  publisher = {American Physical Society},
  doi = {10.1103/PhysRevA.102.052414},
  url = {https://link.aps.org/doi/10.1103/PhysRevA.102.052414}
}

@article{harrigan2021quantum,
  title={Quantum approximate optimization of non-planar graph problems on a planar superconducting processor},
  author={Harrigan, Matthew P and Sung, Kevin J and Neeley, Matthew and Satzinger, Kevin J and Arute, Frank and Arya, Kunal and Atalaya, Juan and Bardin, Joseph C and Barends, Rami and Boixo, Sergio and others},
  journal={Nat. Phys.},
  volume={17},
  number={3},
  pages={332--336},
  year={2021},
  publisher={Nature Publishing Group UK London},
  doi = {10.1038/s41567-020-01105-y},  
}

@article{guerreschi2019qaoa,
  title={{QAOA for Max-Cut requires hundreds of qubits for quantum speed-up}},
  author={Guerreschi, Gian Giacomo and Matsuura, Anne Y},
  journal={Sci Rep},
  volume={9},
  number={1},
  pages={1--7},
  year={2019},
  publisher={Nature Publishing Group},
  doi = {https://doi.org/10.1038/s41598-019-43176-9},
}

@article{sharma2020noise,
  title = {Noise Resilience of Variational Quantum Compiling},
  author = {Sharma, Kunal and Khatri, Sumeet and Cerezo, M and Coles, Patrick J},
  year = {2020},
  month = apr,
  journal = {New Journal of Physics},
  volume = {22},
  number = {4},
  pages = {043006},
  issn = {1367-2630},
  doi = {10.1088/1367-2630/ab784c}
}

@article{cincio2021machine,
  title = {{Machine Learning of Noise-Resilient Quantum Circuits}},
  author={Cincio, Lukasz and Rudinger, Kenneth and Sarovar, Mohan and Coles, Patrick J},
  journal={PRX Quantum},
  volume={2},
  number={1},
  pages={010324},
  year={2021},
  publisher = {American Physical Society},
  doi = {10.1103/PRXQuantum.2.010324},
  url = {https://link.aps.org/doi/10.1103/PRXQuantum.2.010324}
}

@article{panchenko2012sherrington,
  title={{The Sherrington-Kirkpatrick model: an overview}},
  author={Panchenko, Dmitry},
  journal={J. Stat. Phys.},
  volume={149},
  number={2},
  pages={362--383},
  year={2012},
  publisher={Springer},
  doi = {10.1007/s10955-012-0586-7}
}

@article{sherrington1975solvable,
  title = {Solvable Model of a Spin-Glass},
  author = {Sherrington, David and Kirkpatrick, Scott},
  journal = {Phys. Rev. Lett.},
  volume = {35},
  issue = {26},
  pages = {1792--1796},
  numpages = {0},
  year = {1975},
  month = {Dec},
  publisher = {American Physical Society},
  doi = {10.1103/PhysRevLett.35.1792},
  url = {https://link.aps.org/doi/10.1103/PhysRevLett.35.1792}
}

@Article{maier2019environment,
  author    = {Maier, Christine and Brydges, Tiff and Jurcevic, Petar and Trautmann, Nils and Hempel, Cornelius and Lanyon, Ben P. and Hauke, Philipp and Blatt, Rainer and Roos, Christian F.},
  journal   = {Phys. Rev. Lett.},
  title     = {Environment-Assisted Quantum Transport in a 10-qubit Network},
  year      = {2019},
  month     = {Feb},
  pages     = {050501},
  volume    = {122},
  doi       = {10.1103/PhysRevLett.122.050501},
  issue     = {5},
  numpages  = {6},
  publisher = {American Physical Society},
  url       = {https://link.aps.org/doi/10.1103/PhysRevLett.122.050501},
}

@article{smith2016many,
  author    = {Smith, J. and Lee, A. and Richerme, P. and Neyenhuis, B. and Hess, P. W. and Hauke, P. and Heyl, M. and Huse, D. A. and Monroe, C.},
  title     = {Many-body localization in a quantum simulator with programmable random disorder},
  year      = {2016},
  month     = {jun},
  number    = {10},
  pages     = {907--911},
  volume    = {12},
  doi       = {10.1038/nphys3783},
  publisher = {Springer Science and Business Media {LLC}},
  journal = {Nat. Phys.},
}

@article{rabinovich2022ion,
  title={Ion-native variational ansatz for quantum approximate optimization},
  author={Rabinovich, D. and Adhikary, S. and Campos, E. and Akshay, V. and Anikin, E. and Sengupta, R. and Lakhmanskaya, O. and Lakhmanskiy, K. and Biamonte, J.},
  journal={Phys. Rev. A},
  volume={106},
  number={3},
  pages={032418},
  year={2022},
  publisher={APS},
  doi = {10.1103/PhysRevA.106.032418},
  url = {https://link.aps.org/doi/10.1103/PhysRevA.106.032418}
}

@article{weidenfeller2022scaling,
  title={Scaling of the quantum approximate optimization algorithm on superconducting qubit based hardware},
  author={Weidenfeller, Johannes and Valor, Lucia C and Gacon, Julien and Tornow, Caroline and Bello, Luciano and Woerner, Stefan and Egger, Daniel J},
  journal={Quantum},
  volume={6},
  pages={870},
  year={2022},
  publisher={Verein zur F{\"o}rderung des Open Access Publizierens in den Quantenwissenschaften},
  doi = {10.22331/q-2022-12-07-870},
  url = {https://doi.org/10.22331/q-2022-12-07-870},
}

@article{ratcliffe2018scaling,
  title={Scaling trapped ion quantum computers using fast gates and microtraps},
  author={Ratcliffe, Alexander K and Taylor, Richard L and Hope, Joseph J and Carvalho, Andr{\'e} RR},
  journal={Phys. Rev. Lett.},
  volume={120},
  number={22},
  pages={220501},
  year={2018},
  publisher={APS},
  doi = {10.1103/PhysRevLett.120.220501},
  url = {https://link.aps.org/doi/10.1103/PhysRevLett.120.220501}
}

@article{hegde2022toward,
  title={Toward the Speed Limit of High-Fidelity Two-Qubit Gates},
  author={Hegde, Swathi S and Zhang, Jingfu and Suter, Dieter},
  journal={Phys. Rev. Lett.},
  volume={128},
  number={23},
  pages={230502},
  year={2022},
  publisher={APS},
  doi = {10.1103/PhysRevLett.128.230502},
  url = {https://link.aps.org/doi/10.1103/PhysRevLett.128.230502}
}

@article{mills2022two,
  title={Two-qubit silicon quantum processor with operation fidelity exceeding 99\%},
  author={Mills, Adam R and Guinn, Charles R and Gullans, Michael J and Sigillito, Anthony J and Feldman, Mayer M and Nielsen, Erik and Petta, Jason R},
  journal={Sci. Adv.},
  volume={8},
  number={14},
  pages={eabn5130},
  year={2022},
  publisher={American Association for the Advancement of Science},
  doi = {10.1126/sciadv.abn5130},
  URL = {https://www.science.org/doi/abs/10.1126/sciadv.abn5130},
}

@article{kandala2017hardware,
  title={Hardware-efficient variational quantum eigensolver for small molecules and quantum magnets},
  author={Kandala, Abhinav and Mezzacapo, Antonio and Temme, Kristan and Takita, Maika and Brink, Markus and Chow, Jerry M and Gambetta, Jay M},
  journal={Nature},
  volume={549},
  number={7671},
  pages={242--246},
  year={2017},
  publisher={Nature Publishing Group},
  doi = {https://doi.org/10.1038/nature23879}
}

@article{pagano2020quantum,
  title={Quantum approximate optimization of the long-range {I}sing model with a trapped-ion quantum simulator},
  author={Pagano, Guido and Bapat, Aniruddha and Becker, Patrick and Collins, Katherine S and De, Arinjoy and Hess, Paul W and Kaplan, Harvey B and Kyprianidis, Antonis and Tan, Wen Lin and Baldwin, Christopher and others},
  journal={Proc. Natl. Acad. Sci.},
  volume={117},
  number={41},
  pages={25396--25401},
  year={2020},
  publisher={National Acad Sciences},
  doi = {10.1073/pnas.2006373117},
}

@article{cerezo2021variational,
  title={Variational quantum algorithms},
  author={Cerezo, Marco and Arrasmith, Andrew and Babbush, Ryan and Benjamin, Simon C and Endo, Suguru and Fujii, Keisuke and McClean, Jarrod R and Mitarai, Kosuke and Yuan, Xiao and Cincio, Lukasz and others},
  journal={Nat. Rev. Phys.},
  volume={3},
  number={9},
  pages={625--644},
  year={2021},
  publisher={Nature Publishing Group},
  doi = {https://doi.org/10.1038/s42254-021-00348-9}
}

@article{biamonte2021universal,
  title={Universal variational quantum computation},
  author={Biamonte, Jacob},
  journal={Phys. Rev. A},
  volume={103},
  number={3},
  pages={L030401},
  year={2021},
  publisher = {American Physical Society},
  doi = {10.1103/PhysRevA.103.L030401},
  url = {https://link.aps.org/doi/10.1103/PhysRevA.103.L030401}
}

@misc{verdon2017quantum,
      title={{A quantum algorithm to train neural networks using low-depth circuits}}, 
      author={Verdon, Guillaume and Broughton, Michael and Biamonte, Jacob},
      year={2017},
      eprint={1712.05304},
      archivePrefix={arXiv},
      primaryClass={quant-ph},
      url={https://arxiv.org/abs/1712.05304}, 
}

@article{liang2020variational,
  title={Variational quantum algorithms for dimensionality reduction and classification},
  author={Liang, Jin-Min and Shen, Shu-Qian and Li, Ming and Li, Lei},
  journal={Phys. Rev. A},
  volume={101},
  number={3},
  pages={032323},
  year={2020},
  publisher = {American Physical Society},
  doi = {10.1103/PhysRevA.101.032323},
  url = {https://link.aps.org/doi/10.1103/PhysRevA.101.032323}
}

@article{mcclean2016theory,
  title={The theory of variational hybrid quantum-classical algorithms},
  author={McClean, Jarrod R and Romero, Jonathan and Babbush, Ryan and Aspuru-Guzik, Al{\'a}n},
  journal={New J. Phys.},
  volume={18},
  number={2},
  pages={023023},
  year={2016},
  publisher={IOP Publishing},
  doi = {10.1088/1367-2630/18/2/023023},
  url = {https://dx.doi.org/10.1088/1367-2630/18/2/023023},
}

@article{benedetti2019parameterized,
  title={Parameterized quantum circuits as machine learning models},
  author={Benedetti, Marcello and Lloyd, Erika and Sack, Stefan and Fiorentini, Mattia},
  journal={Quantum Sci. Technol.},
  volume={4},
  number={4},
  pages={043001},
  year={2019},
  publisher={IOP Publishing},
  doi = {10.1088/2058-9565/ab4eb5},
  url = {https://dx.doi.org/10.1088/2058-9565/ab4eb5},
}

@article{yuan2019theory,
  title={Theory of variational quantum simulation},
  author={Yuan, Xiao and Endo, Suguru and Zhao, Qi and Li, Ying and Benjamin, Simon C},
  journal={Quantum},
  volume={3},
  pages={191},
  year={2019},
  publisher={Verein zur F{\"o}rderung des Open Access Publizierens in den Quantenwissenschaften},
  doi = {10.22331/q-2019-10-07-191}
}

@article{cirstoiu2020variational,
  title={Variational fast forwarding for quantum simulation beyond the coherence time},
  author={Cirstoiu, Cristina and Holmes, Zoe and Iosue, Joseph and Cincio, Lukasz and Coles, Patrick J and Sornborger, Andrew},
  journal={npj Quantum Inf},
  volume={6},
  number={1},
  pages={82},
  year={2020},
  publisher={Nature Publishing Group UK London},
  doi = {10.1038/s41534-020-00302-0}
}

@article{gibbs2021long,
	author = {Gibbs, Joe and Gili, Kaitlin and Holmes, Zo{\"e} and Commeau, Benjamin and Arrasmith, Andrew and Cincio, Lukasz and Coles, Patrick J. and Sornborger, Andrew},
	doi = {10.1038/s41534-022-00625-0},
	isbn = {2056-6387},
	journal = {npj Quant Inf},
	number = {1},
	pages = {135},
	title = {Long-time simulations for fixed input states on quantum hardware},
	url = {https://doi.org/10.1038/s41534-022-00625-0},
	volume = {8},
	year = {2022},
}

@article{endo2020variational,
  title={Variational quantum simulation of general processes},
  author={Endo, Suguru and Sun, Jinzhao and Li, Ying and Benjamin, Simon C and Yuan, Xiao},
  journal={Phys. Rev. Lett.},
  volume={125},
  number={1},
  pages={010501},
  year={2020},
  publisher = {American Physical Society},
  doi = {10.1103/PhysRevLett.125.010501},
  url = {https://link.aps.org/doi/10.1103/PhysRevLett.125.010501}
}

@article{jones2022robust,
  title={Robust quantum compilation and circuit optimisation via energy minimisation},
  author={Jones, Tyson and Benjamin, Simon C},
  journal={Quantum},
  volume={6},
  pages={628},
  year={2022},
  publisher={Verein zur F{\"o}rderung des Open Access Publizierens in den Quantenwissenschaften},
  doi = {10.22331/q-2022-01-24-628}
}

@article{khatri2019quantum,
  doi = {10.22331/q-2019-05-13-140},
  url = {https://doi.org/10.22331/q-2019-05-13-140},
  title = {Quantum-assisted quantum compiling},
  author = {Khatri, Sumeet and LaRose, Ryan and Poremba, Alexander and Cincio, Lukasz and Sornborger, Andrew T. and Coles, Patrick J.},
  journal = {{Quantum}},
  issn = {2521-327X},
  publisher = {{Verein zur F{\"{o}}rderung des Open Access Publizierens in den Quantenwissenschaften}},
  volume = {3},
  pages = {140},
  month = may,
  year = {2019}
}

@article{mitarai2018quantum,
  title={Quantum circuit learning},
  author={Mitarai, Kosuke and Negoro, Makoto and Kitagawa, Masahiro and Fujii, Keisuke},
  journal={Phys. Rev. A},
  volume={98},
  number={3},
  pages={032309},
  year={2018},
  publisher = {American Physical Society},
  doi = {10.1103/PhysRevA.98.032309},
  url = {https://link.aps.org/doi/10.1103/PhysRevA.98.032309}
}

@article{schuld2019evaluating,
  title={Evaluating analytic gradients on quantum hardware},
  author={Schuld, Maria and Bergholm, Ville and Gogolin, Christian and Izaac, Josh and Killoran, Nathan},
  journal={Phys. Rev. A},
  volume={99},
  number={3},
  pages={032331},
  year={2019},
  publisher = {American Physical Society},
  doi = {10.1103/PhysRevA.99.032331},
  url = {https://link.aps.org/doi/10.1103/PhysRevA.99.032331}
}

@article{xu2019variational,
  title = {{Variational Circuit Compiler for Quantum Error Correction}},
  author = {Xu, Xiaosi and Benjamin, Simon C. and Yuan, Xiao},
  journal = {Phys. Rev. Appl.},
  volume = {15},
  issue = {3},
  pages = {034068},
  numpages = {13},
  year = {2021},
  month = {Mar},
  publisher = {American Physical Society},
  doi = {10.1103/PhysRevApplied.15.034068},
  url = {https://link.aps.org/doi/10.1103/PhysRevApplied.15.034068}
}

@article{cerezo2021cost,
  title={Cost function dependent barren plateaus in shallow parametrized quantum circuits},
  author={Cerezo, Marco and Sone, Akira and Volkoff, Tyler and Cincio, Lukasz and Coles, Patrick J},
  journal={Nat. Commun.},
  volume={12},
  number={1},
  pages={1791},
  year={2021},
  publisher={Nature Publishing Group UK London},
  doi = {https://doi.org/10.1038/s41467-021-21728-w},
}

@article{uvarov2020frustrated,
  title = {Variational quantum eigensolver for frustrated quantum systems},
  author = {Uvarov, Alexey and Biamonte, Jacob D. and Yudin, Dmitry},
  journal = {Phys. Rev. B},
  volume = {102},
  issue = {7},
  pages = {075104},
  numpages = {6},
  year = {2020},
  month = {Aug},
  publisher = {American Physical Society},
  doi = {10.1103/PhysRevB.102.075104},
  url = {https://link.aps.org/doi/10.1103/PhysRevB.102.075104}
}

@article{parrish2019quantum,
  title={Quantum computation of electronic transitions using a variational quantum eigensolver},
  author={Parrish, Robert M and Hohenstein, Edward G and McMahon, Peter L and Mart{\'\i}nez, Todd J},
  journal={Phys. Rev. Lett.},
  volume={122},
  number={23},
  pages={230401},
  year={2019},
  publisher = {American Physical Society},
  doi = {10.1103/PhysRevLett.122.230401},
  url = {https://link.aps.org/doi/10.1103/PhysRevLett.122.230401}
}

@article{peruzzo2014variational,
  title={A variational eigenvalue solver on a photonic quantum processor},
  author={Peruzzo, Alberto and McClean, Jarrod and Shadbolt, Peter and Yung, Man-Hong and Zhou, Xiao-Qi and Love, Peter J and Aspuru-Guzik, Al{\'a}n and O’Brien, Jeremy L},
  journal={Nat. Commun.},
  volume={5},
  pages={4213},
  year={2014},
  publisher={Nature Publishing Group},
  doi = {https://doi.org/10.1038/ncomms5213}
}

@article{zeng2021simulating,
  title={Simulating noisy variational quantum eigensolver with local noise models},
  author={Zeng, Jinfeng and Wu, Zipeng and Cao, Chenfeng and Zhang, Chao and Hou, Shi-Yao and Xu, Pengxiang and Zeng, Bei},
  journal={Quantum Engineering},
  volume={3},
  number={4},
  pages={e77},
  year={2021},
  publisher={Wiley Online Library},
  doi = {10.1002/que2.77},
  url = {https://doi.org/10.1002/que2.77}
}

@article{kardashin2021numerical,
  title={Numerical hardware-efficient variational quantum simulation of a soliton solution},
  author={Kardashin, A. and Pervishko, A. and Biamonte, J. and Yudin, D.},
  journal={Phys. Rev. A},
  volume={104},
  number={2},
  pages={L020402},
  year={2021},
  publisher = {American Physical Society},
  doi = {10.1103/PhysRevA.104.L020402},
  url = {https://link.aps.org/doi/10.1103/PhysRevA.104.L020402}
}

@article{uvarov2024mitigating,
  title = {Mitigating quantum gate errors for variational eigensolvers using hardware-inspired zero-noise extrapolation},
  author = {Uvarov, Alexey and Rabinovich, Daniil and Lakhmanskaya, Olga and Lakhmanskiy, Kirill and Biamonte, Jacob and Adhikary, Soumik},
  journal = {Phys. Rev. A},
  volume = {110},
  issue = {1},
  pages = {012404},
  numpages = {11},
  year = {2024},
  month = {Jul},
  publisher = {American Physical Society},
  doi = {10.1103/PhysRevA.110.012404},
  url = {https://link.aps.org/doi/10.1103/PhysRevA.110.012404}
}

@Article{zhang2017observation,
  author   = {Zhang, J. and Pagano, G. and Hess, P. W. and Kyprianidis, A. and Becker, P. and Kaplan, H. and Gorshkov, A. V. and Gong, Z.-X. and Monroe, C.},
  journal  = {Nature},
  title    = {Observation of a many-body dynamical phase transition with a 53-qubit quantum simulator},
  year     = {2017},
  issn     = {1476-4687},
  number   = {7682},
  pages    = {601--604},
  volume   = {551},
  doi      = {10.1038/nature24654},
  refid    = {Zhang2017},
  url      = {https://doi.org/10.1038/nature24654},
}

@Article{pogorelov2021compact,
  author    = {Pogorelov, I. and Feldker, T. and Marciniak, Ch. D. and Postler, L. and Jacob, G. and Krieglsteiner, O. and Podlesnic, V. and Meth, M. and Negnevitsky, V. and Stadler, M. and H\"ofer, B. and W\"achter, C. and Lakhmanskiy, K. and Blatt, R. and Schindler, P. and Monz, T.},
  journal   = {PRX Quantum},
  title     = {Compact Ion-Trap Quantum Computing Demonstrator},
  year      = {2021},
  month     = {Jun},
  pages     = {020343},
  volume    = {2},
  doi       = {10.1103/PRXQuantum.2.020343},
  issue     = {2},
  numpages  = {23},
  publisher = {American Physical Society},
  url       = {https://link.aps.org/doi/10.1103/PRXQuantum.2.020343},
}

@Article{wright2019benchmarking,
  author   = {Wright, K. and Beck, K. M. and Debnath, S. and Amini, J. M. and Nam, Y. and Grzesiak, N. and Chen, J.-S. and Pisenti, N. C. and Chmielewski, M. and Collins, C. and Hudek, K. M. and Mizrahi, J. and Wong-Campos, J. D. and Allen, S. and Apisdorf, J. and Solomon, P. and Williams, M. and Ducore, A. M. and Blinov, A. and Kreikemeier, S. M. and Chaplin, V. and Keesan, M. and Monroe, C. and Kim, J.},
  journal  = {Nature Communications},
  title    = {Benchmarking an 11-qubit quantum computer},
  year     = {2019},
  issn     = {2041-1723},
  number   = {1},
  pages    = {5464},
  volume   = {10},
  doi      = {10.1038/s41467-019-13534-2},
  refid    = {Wright2019},
  url      = {https://doi.org/10.1038/s41467-019-13534-2},
}

@article{rabinovich2024robustness,
  title = {Robustness of variational quantum algorithms against stochastic parameter perturbation},
  author = {Rabinovich, Daniil and Campos, Ernesto and Adhikary, Soumik and Pankovets, Ekaterina and Vinichenko, Dmitry and Biamonte, Jacob},
  journal = {Phys. Rev. A},
  volume = {109},
  issue = {4},
  pages = {042426},
  numpages = {9},
  year = {2024},
  month = {Apr},
  publisher = {American Physical Society},
  doi = {10.1103/PhysRevA.109.042426},
  url = {https://link.aps.org/doi/10.1103/PhysRevA.109.042426}
}

@article{cade2020strategies,
  title={Strategies for solving the {F}ermi-{H}ubbard model on near-term quantum computers},
  author={Cade, Chris and Mineh, Lana and Montanaro, Ashley and Stanisic, Stasja},
  journal={Phys. Rev. B},
  volume={102},
  number={23},
  pages={235122},
  year={2020},
  publisher={APS},
  doi = {10.1103/PhysRevB.102.235122},
  url = {https://link.aps.org/doi/10.1103/PhysRevB.102.235122}
}

@article{hempel2018quantum,
title = {Quantum Chemistry Calculations on a Trapped-Ion Quantum Simulator},
  author = {Hempel, Cornelius and Maier, Christine and Romero, Jonathan and McClean, Jarrod and Monz, Thomas and Shen, Heng and Jurcevic, Petar and Lanyon, Ben P. and Love, Peter and Babbush, Ryan and Aspuru-Guzik, Al\'an and Blatt, Rainer and Roos, Christian F.},
  journal = {Phys. Rev. X},
  volume = {8},
  issue = {3},
  pages = {031022},
  numpages = {22},
  year = {2018},
  month = {Jul},
  publisher = {American Physical Society},
  doi = {10.1103/PhysRevX.8.031022},
  url = {https://link.aps.org/doi/10.1103/PhysRevX.8.031022}
}

@article{he2021variational,
  title={{Variational quantum compiling with double Q-learning}},
  author={He, Zhimin and Li, Lvzhou and Zheng, Shenggen and Li, Yongyao and Situ, Haozhen},
  journal={New J. Phys.},
  volume={23},
  number={3},
  pages={033002},
  year={2021},
  publisher={IOP Publishing},
  doi = {10.1088/1367-2630/abe0ae},
  url = {https://dx.doi.org/10.1088/1367-2630/abe0ae},
}

@article{AHC2022,
doi = {10.1088/2058-9565/ac7d06},
url = {https://dx.doi.org/10.1088/2058-9565/ac7d06},
year = {2022},
month = {aug},
publisher = {IOP Publishing},
volume = {7},
number = {4},
pages = {045015},
author = {Arrasmith, A. and Holmes, Z. and Cerezo, M. and Coles, P. J.},
title = {Equivalence of quantum barren plateaus to cost concentration and narrow gorges},
journal = {Quantum Sci. Technol.},
}

@article{AK2022,
	author = {Anschuetz, E. R. and Kiani, B. T.},
	date = {2022/12/15},
	doi = {10.1038/s41467-022-35364-5},
	id = {Anschuetz2022},
	isbn = {2041-1723},
	journal = {Nat. Commun.},
	number = {1},
	pages = {7760},
	title = {Quantum variational algorithms are swamped with traps},
	url = {https://doi.org/10.1038/s41467-022-35364-5},
	volume = {13},
	year = {2022},
}

@article{CSV2021,
	author = {Cerezo, M. and Sone, A. and Volkoff, T. and Cincio, L. and Coles, P. J.},
	date = {2021/03/19},
	doi = {10.1038/s41467-021-21728-w},
	id = {Cerezo2021},
	isbn = {2041-1723},
	journal = {Nat. Commun.},
	number = {1},
	pages = {1791},
	title = {Cost function dependent barren plateaus in shallow parametrized quantum circuits},
	url = {https://doi.org/10.1038/s41467-021-21728-w},
	volume = {12},
	year = {2021},
}

@book{DS96,
  author = {Dennis, J.~E. and Schnabel, R.~B.},
  title = {Numerical Methods for Unconstrained Optimization and Nonlinear Equations},
  publisher = {SIAM},
  year = {1996},
  address = {Philadelphia},
  doi = {https://doi.org/10.1137/1.9781611971200},
}

@misc{Farhi2014,
      title={{A Quantum Approximate Optimization Algorithm}}, 
      author={Farhi, E. and Goldstone, J. and Gutmann, S.},
      year={2014},
      eprint={1411.4028},
      archivePrefix={arXiv},
      primaryClass={quant-ph},
      url={https://arxiv.org/abs/1411.4028}, 
}

@article{Farhi2022,
  doi = {10.22331/q-2022-07-07-759},
  url = {https://doi.org/10.22331/q-2022-07-07-759},
  title = {The {Q}uantum {A}pproximate {O}ptimization {A}lgorithm and the {S}herrington-{K}irkpatrick {M}odel at {I}nfinite {S}ize},
  author = {Farhi, E. and Goldstone, J. and Gutmann, S. and Zhou, L.},
  journal = {{Quantum}},
  issn = {2521-327X},
  publisher = {{Verein zur F{\"{o}}rderung des Open Access Publizierens in den Quantenwissenschaften}},
  volume = {6},
  pages = {759},
  month = jul,
  year = {2022}
}

@article{holmes2022connecting,
  title = {{Connecting Ansatz Expressibility to Gradient Magnitudes and Barren Plateaus}},
  author = {Holmes, Z. and Sharma, K. and Cerezo, M. and Coles, P. J.},
  journal = {PRX Quantum},
  volume = {3},
  issue = {1},
  pages = {010313},
  numpages = {23},
  year = {2022},
  month = {Jan},
  publisher = {American Physical Society},
  doi = {10.1103/PRXQuantum.3.010313},
  url = {https://link.aps.org/doi/10.1103/PRXQuantum.3.010313}
}

@article{James1998,
	author = {James, D. F. V.},
	doi = {10.1007/s003400050373},
	journal = {Appl. Phys. B},
	number = {2},
	pages = {181--190},
	title = {Quantum dynamics of cold trapped ions with application to quantum computation},
	url = {https://doi.org/10.1007/s003400050373},
	volume = {66},
	year = {1998},
}

@article{Kim2009,
  title = {{Entanglement and Tunable Spin-Spin Couplings between Trapped Ions Using Multiple Transverse Modes}},
  author = {Kim, K. and Chang, M.-S. and Islam, R. and Korenblit, S. and Duan, L.-M. and Monroe, C.},
  journal = {Phys. Rev. Lett.},
  volume = {103},
  issue = {12},
  pages = {120502},
  numpages = {4},
  year = {2009},
  month = {Sep},
  publisher = {American Physical Society},
  doi = {10.1103/PhysRevLett.103.120502},
  url = {https://link.aps.org/doi/10.1103/PhysRevLett.103.120502}
}

@article{Lee2005,
doi = {10.1088/1464-4266/7/10/025},
url = {https://dx.doi.org/10.1088/1464-4266/7/10/025},
year = {2005},
month = {sep},
publisher = {},
volume = {7},
number = {10},
pages = {S371},
author = {Lee, P. J. and Brickman, K.-A. and Deslauriers, L. and Haljan, P. C and Duan, L.-M. and Monroe, C.},
title = {Phase control of trapped ion quantum gates},
journal = {J. Opt. B Quantum Semiclassical Opt.},
}

@article{monroe2021programmable,
  title = {Programmable quantum simulations of spin systems with trapped ions},
  author = {Monroe, C. and Campbell, W. C. and Duan, L.-M. and Gong, Z.-X. and Gorshkov, A. V. and Hess, P. W. and Islam, R. and Kim, K. and Linke, N. M. and Pagano, G. and Richerme, P. and Senko, C. and Yao, N. Y.},
  journal = {Rev. Mod. Phys.},
  volume = {93},
  issue = {2},
  pages = {025001},
  numpages = {57},
  year = {2021},
  month = {Apr},
  publisher = {American Physical Society},
  doi = {10.1103/RevModPhys.93.025001},
  url = {https://link.aps.org/doi/10.1103/RevModPhys.93.025001}
}

@article{Powell1964,
    author = {Powell, M. J. D.},
    title = {An efficient method for finding the minimum of a function of several variables without calculating derivatives},
    journal = {Comput. J.},
    volume = {7},
    number = {2},
    pages = {155-162},
    year = {1964},
    month = {01},
    issn = {0010-4620},
    doi = {10.1093/comjnl/7.2.155},
    url = {https://doi.org/10.1093/comjnl/7.2.155},
}

@Inbook{Powell1994,
    author = {Powell, M. J. D.},
    editor = {Gomez, S. and Hennart, J.-P.},
    title = {A Direct Search Optimization Method That Models the Objective and Constraint Functions by Linear Interpolation},
    bookTitle = {Advances in Optimization and Numerical Analysis},
    year = {1994},
    publisher = {Springer},
    address = {Dordrecht},
    pages = {51--67},
    doi = {10.1007/978-94-015-8330-5_4},
    url = {https://doi.org/10.1007/978-94-015-8330-5_4},
}

@article{headley2022approximating,
  title={Approximating the quantum approximate optimization algorithm with digital-analog interactions},
  author={Headley, David and M{\"u}ller, Thorge and Martin, Ana and Solano, Enrique and Sanz, Mikel and Wilhelm, Frank K},
  journal={Physical Review A},
  volume={106},
  number={4},
  pages={042446},
  year={2022},
  publisher={APS}
}

@article{gonzalez2021digital,
  title={Digital-analog quantum simulations using the cross-resonance effect},
  author={Gonzalez-Raya, Tasio and Asensio-Perea, Rodrigo and Martin, Ana and C{\'e}leri, Lucas C and Sanz, Mikel and Lougovski, Pavel and Dumitrescu, Eugene F},
  journal={PRX Quantum},
  volume={2},
  number={2},
  pages={020328},
  year={2021},
  publisher={APS}
}

@article{martin2020digital,
  title={Digital-analog quantum algorithm for the quantum Fourier transform},
  author={Martin, Ana and Lamata, Lucas and Solano, Enrique and Sanz, Mikel},
  journal={Physical Review Research},
  volume={2},
  number={1},
  pages={013012},
  year={2020},
  publisher={APS}
}

@article{garcia2024digital,
  title={Digital-analog quantum computation with arbitrary two-body Hamiltonians},
  author={Garcia-de-Andoin, Mikel and Saiz, {\'A}lvaro and P{\'e}rez-Fern{\'a}ndez, Pedro and Lamata, Lucas and Oregi, Izaskun and Sanz, Mikel},
  journal={Physical Review Research},
  volume={6},
  number={1},
  pages={013280},
  year={2024},
  publisher={APS}
}

@article{yu2022superconducting,
  title={Superconducting circuit architecture for digital-analog quantum computing},
  author={Yu, Jing and Retamal, Juan Carlos and Sanz, Mikel and Solano, Enrique and Albarr{\'a}n-Arriagada, Francisco},
  journal={EPJ Quantum Technology},
  volume={9},
  number={1},
  pages={1--35},
  year={2022},
  publisher={Springer}
}

@article{parra2020digital,
  title={Digital-analog quantum computation},
  author={Parra-Rodriguez, Adrian and Lougovski, Pavel and Lamata, Lucas and Solano, Enrique and Sanz, Mikel},
  journal={Physical Review A},
  volume={101},
  number={2},
  pages={022305},
  year={2020},
  publisher={APS}
}

@article{PPY2024,
  title = {Probabilistic tensor optimization of quantum circuits for the $\text{max}\text{\ensuremath{-}}k\text{\ensuremath{-}}\text{cut}$ problem},
  author = {Paradezhenko, G. V. and Pervishko, A. A. and Yudin, D.},
  journal = {Phys. Rev. A},
  volume = {109},
  issue = {1},
  pages = {012436},
  numpages = {8},
  year = {2024},
  month = {Jan},
  publisher = {American Physical Society},
  doi = {10.1103/PhysRevA.109.012436},
  url = {https://link.aps.org/doi/10.1103/PhysRevA.109.012436}
}

@inproceedings{RGB2024, 
    author = {Rajakumar, J. and Golden, J. and B\"{a}rtschi, A. and Eidenbenz, S.}, 
    title = {{Trainability Barriers in Low-Depth QAOA Landscapes}}, 
    year = {2024}, 
    isbn = {9798400705977}, 
    publisher = {Association for Computing Machinery}, 
    address = {New York}, 
    url = {https://doi.org/10.1145/3649153.3649204}, 
    doi = {10.1145/3649153.3649204}, 
    booktitle = {Proceedings of the 21st ACM International Conference on Computing Frontiers}, 
    pages = {199–206}, 
    numpages = {8}, 
    location = {Ischia, Italy}, 
    series = {CF '24}
}

@article{SciPy,
  author  = {Virtanen, P. and et al.},
  title   = {{{SciPy} 1.0: Fundamental Algorithms for Scientific
            Computing in Python}},
  journal = {Nature Methods},
  year    = {2020},
  volume  = {17},
  pages   = {261--272},
  adsurl  = {https://rdcu.be/b08Wh},
  doi     = {10.1038/s41592-019-0686-2},
}

@article{SM2000,
  title = {Entanglement and quantum computation with ions in thermal motion},
  author = {S\o{}rensen, A. and M\o{}lmer, K.},
  journal = {Phys. Rev. A},
  volume = {62},
  issue = {2},
  pages = {022311},
  numpages = {11},
  year = {2000},
  month = {Jul},
  publisher = {American Physical Society},
  doi = {10.1103/PhysRevA.62.022311},
  url = {https://link.aps.org/doi/10.1103/PhysRevA.62.022311}
}

@article{SJA2019,
    author = {Sim, S. and Johnson, P. D. and Aspuru-Guzik, A.},
    title = {{Expressibility and Entangling Capability of Parameterized Quantum Circuits for Hybrid Quantum-Classical Algorithms}},
    journal = {Adv. Quantum Technol.},
    volume = {2},
    number = {12},
    pages = {1900070},
    doi = {https://doi.org/10.1002/qute.201900070},
    year = {2019}
}

@article{Tseng2001,
	author = {Tseng, P.},
	date = {2001/06/01},
	doi = {10.1023/A:1017501703105},
	id = {Tseng2001},
	isbn = {1573-2878},
	journal = {J. Optim. Theory Appl.},
	number = {3},
	pages = {475--494},
	title = {{Convergence of a Block Coordinate Descent Method for Nondifferentiable Minimization}},
	url = {https://doi.org/10.1023/A:1017501703105},
	volume = {109},
	year = {2001}
}

@article{Zhores,
  author = {Zacharov, I. and Arslanov, R. and Gunin, M. and Stefonishin, D. and Bykov, A. and Pavlov, S. and Panarin, O. and Maliutin, A. and Rykovanov, S. and Fedorov, M.},
  doi = {doi:10.1515/eng-2019-0059},
  url = {https://doi.org/10.1515/eng-2019-0059},
  title = {``{Z}hores'' -- {P}etaflops supercomputer for data-driven modeling, machine learning and artificial intelligence installed in {S}kolkovo {I}nstitute of {S}cience and {T}echnology},
  journal = {Open Eng.},
  number = {1},
  volume = {9},
  year = {2019},
  pages = {512--520}
}

@article{ZMD2006,
  title = {{Trapped Ion Quantum Computation with Transverse Phonon Modes}},
  author = {Zhu, S.-L. and Monroe, C. and Duan, L.-M.},
  journal = {Phys. Rev. Lett.},
  volume = {97},
  issue = {5},
  pages = {050505},
  numpages = {4},
  year = {2006},
  month = {Aug},
  publisher = {American Physical Society},
  doi = {10.1103/PhysRevLett.97.050505},
  url = {https://link.aps.org/doi/10.1103/PhysRevLett.97.050505}
}

@article{ZMD2006a,
doi = {10.1209/epl/i2005-10424-4},
url = {https://dx.doi.org/10.1209/epl/i2005-10424-4},
year = {2006},
month = {jan},
publisher = {},
volume = {73},
number = {4},
pages = {485},
author = {Zhu, S.-L. and Monroe, C. and Duan, L.-M.},
title = {Arbitrary-speed quantum gates within large ion crystals  through minimum control of laser beams},
journal = {Europhys. Lett.},
}

@article{Anikin2025fast,
  title = {Fast Molmer-S\o{}rensen gates in trapped-ion quantum processors with compensated carrier transition},
  author = {Anikin, Evgeny and Chuchalin, Andrey and Morozov, Nikita and Lakhmanskaya, Olga and Lakhmanskiy, Kirill},
  journal = {Phys. Rev. Appl.},
  volume = {24},
  issue = {1},
  pages = {014044},
  numpages = {15},
  year = {2025},
  month = {Jul},
  publisher = {American Physical Society},
  doi = {10.1103/kbwc-t19c},
  url = {https://link.aps.org/doi/10.1103/kbwc-t19c}
}

@article{Bharti2022noisy,
  title = {Noisy intermediate-scale quantum algorithms},
  author = {Bharti, Kishor and Cervera-Lierta, Alba and Kyaw, Thi Ha and Haug, Tobias and Alperin-Lea, Sumner and Anand, Abhinav and Degroote, Matthias and Heimonen, Hermanni and Kottmann, Jakob S. and Menke, Tim and Mok, Wai-Keong and Sim, Sukin and Kwek, Leong-Chuan and Aspuru-Guzik, Al\'an},
  journal = {Rev. Mod. Phys.},
  volume = {94},
  issue = {1},
  pages = {015004},
  numpages = {69},
  year = {2022},
  month = {Feb},
  publisher = {American Physical Society},
  doi = {10.1103/RevModPhys.94.015004},
  url = {https://link.aps.org/doi/10.1103/RevModPhys.94.015004}
}

@software{github,
author = {Paradezhenko, Georgii and Rabinovich, Daniil and Campos, Ernesto and Lakhmanskiy, Kirill},
title = {{Heuristic for searching problem specific hyperparameters of ion native ansatz}},
url = {https://github.com/pargv/heuristic_trapped_ions/},
version = {1.0}
}

\appendix

\section{Effective Ising Hamiltonian for laser-ions interaction}
\label{appendix:ions}

A chain of $n$ ions in a linear trap interacting with a laser field of frequency $\omega$ is described by the Hamiltonian $H = H_0 + H_{\rm I}$~\cite{SM2000}, where (here and hereafter, we set~$\hbar \equiv~1$)
\begin{eqnarray}
    H_0 & = & \sum_{m=1}^n \omega_m^{} \left( a_{m}^{\dagger} a_{m}^{} + 1/2 \right) + \frac{\omega_0}{2} \sum_{i=1}^n Z_i, \label{H0-def} \\
    H_{\rm I} & = & \sum_{i=1}^n \Omega_i^{} X_i \cos(k x_i - \omega t + \varphi). \label{HI-def}
\end{eqnarray}
The first term in $H_0$ in Eq.~\eqref{H0-def} describes the ions motion in terms of the normal mode phonon creation and annihilation operators $a_m^{\dagger}$ and $a_m^{}$, respectivelly, at frequency $\omega_m$. The second term in Eq.~\eqref{H0-def} describes the electronic state of isolated individual ions, where $\omega_0 = E_1 - E_0$ is the energy
difference between the excited and ground states. 
The Hamiltonian $H_{\rm I}$ in Eq.~\eqref{HI-def} corresponds to the laser-ion interaction, where $k$ and $\varphi$ are the wave number and phase of a laser beam, while $\Omega_i$ and $x_i$ are the Rabi frequency and position of the $i$-th ion. The ions positions are expressed in terms of phonon operators as
\begin{equation}\label{ions-positions}
    k \cdot x_i 
    = \sum_m \eta_{i}^m \left(a^{}_m e^{-i\omega_m t} + a^{\dagger}_m e^{i\omega_m t}\right),
\end{equation}
where $\eta_i^m$ is the Lamb-Dicke parameter for the coupling between the $i$-th ion and $m$-th normal mode. In Eqs.~\eqref{H0-def} and \eqref{HI-def}, $Z_i$ and $X_i$ denote the Pauli operators acting on the electronic state of the $i$-th ion. 

A standard approach for implementing two-qubit operations in this setting consists of using a pair of noncopropagating laser beams with
bichromatic beatnotes at frequencies $\omega = \omega_0 \pm \mu$ symmetrically detuned by~$\mu$~\cite{Lee2005}. We assume that their wave vectors difference $\bm{\delta k}$ is aligned along the $x$ direction of ions motion. Then, under the rotating wave approximation ($\omega_0 \gg \mu$) and within the Lamb-Dicke limit ($\delta k \langle x_i \rangle \ll 1$), the laser-ion Hamiltonian $H_{\rm I}$ in the interaction picture with respect to $H_0$ reduces to~\cite{Kim2009}
\begin{equation}\label{HI-reduced}
    H_{\mathrm{I}}(t) 
    = \sum_i \Omega_i \left( \delta k \cdot x_i \right) X_i \sin(\mu t). 
\end{equation}

The evolution operator under the Hamiltonian~\eqref{HI-reduced} can be obtained by means of the Magnus expansion terminated after the first two terms~\cite{ZMD2006,ZMD2006a}:
\begin{equation}\label{UI}
     U_{\mathrm{I}}(\tau) 
     = \exp\left[ \sum_i \phi_i(\tau) X_i - i \sum_{i < j} \chi_{ij}(\tau) X_i X_j  \right]. 
\end{equation}
The first term in the exponent in Eq.~\eqref{UI},
\begin{equation}\label{phi}
    \phi_i(\tau) = \sum_m \left[ \alpha_{i,m}^{}(\tau) a_m^{\dagger} - \alpha_{i,m}^{*}(\tau) a_m^{} \right],
\end{equation}
represents spin-dependent displacements of the $m$-th normal mode through phase space by an amount of $\alpha_{i,m}(\tau)$. The second term in Eq.~\eqref{UI} describes a spin-spin interaction
between the $i$-th and $j$-th ions with coupling $\chi_{ij}(\tau)$. 
The explicit expressions for $\alpha_{i,m}(\tau)$ and $\chi_{ij}(\tau)$ are cumbersome, hence we refer to~\cite{ZMD2006}. In what follows, we focus on the dispersive regime ($|\mu - \nu_m| \gg \Omega_i \eta_i^m$), where the phonons are only virtually excited as the displacements become negligible ($\alpha_{i,n}(\tau) \ll 1$)~\cite{ZMD2006}. As a result, the evolution operator~\eqref{UI} takes the form $U_I(\tau) = e^{-i\tau H_{\rm I}}$, where $H_{\rm I}$ reduces to the effective fully-connected Ising Hamiltonian~\eqref{HI-native}~\cite{ZMD2006,monroe2021programmable}.

\section{\label{appendix:phonons}Calculation of Ising couplings in the ion Hamiltonian}

In order to calculate the matrix $C_{ij}$ of phonon contribution to the Ising couplings~\eqref{J-def} in the ion Hamiltonian, we simulated the motion of trapped ions to get the phonon frequencies $\omega_m$ and Lamb-Dicke parameters~$\eta_i^m$ \cite{Anikin2025fast}. This was done using the standard method~\cite{ZMD2006,James1998}.

Let us consider a chain of $n$ ions confined in a
linear trap. The motion of ions is described by the potential
\begin{equation}\label{ions-pot}
    V = \sum_{\xi} \sum_i \frac{M \omega_{\xi} \, \xi_i^2}{2}  + \sum_{i \neq j} \frac{e^2}{4 \pi \epsilon_0 r_{ij}},
\end{equation}
which accounts for the external trapping potential as well as the Coulomb interaction between the ions. In Eq.~\eqref{ions-pot}, $M$ is the ion mass, $\omega_\xi$ is the center-of-mass trap frequency along the direction $\xi = \lbrace x, y, z \rbrace$, and $r_{ij}$ is the distance between the $i$-th and $j$-th ions. Typically, $\omega_x \sim \omega_y \gg \omega_z$, and one has a linear geometry with the chain of ions stretched along the $z$ axis. 
The equilibrium positions $z_i^0$ ($x_i^0 = y_i^0 \equiv 0$) of ions are calculated by minimizing the potential~\eqref{ions-pot} over $z_i$:
\begin{equation}\label{ions-pot-1}
    \tilde{V} = \sum_i u_i^2  + \sum_{i \neq j} \frac{1}{|u_i - u_j|} \to \min_{u_i},
\end{equation}
where $u_i \equiv z_i^0/\ell$ are the dimensionless equilibrium positions, and $\ell = \sqrt[3]{e^2/(4\pi \epsilon_0 M \omega_z^2)}$ sets the scale of ion spacings in the linear chain. 

The phonon frequencies are obtained through diagonalization of the potential $V$ in the harmonic approximation.
Introducing the ions displacements $q^\xi_i = \xi_i - \xi_i^0$, one can expand the potential~\eqref{ions-pot} around the equilibrium positions $\xi^0_i = (x_i^0, y_i^0, z_i^0)$ as
\begin{equation}\label{V-expansion}
    V = \frac{M\omega_z^2}{2} \sum_{\xi} \sum_{ij} A_{ij}^{\xi} q^\xi_i q^\xi_j.
\end{equation}
The matrix elements $A_{ij}^{\xi} = \left[ \partial^2 V / \partial \xi_i \partial \xi_j \right]|_{\xi_i^{} = \xi_i^0}$ in the expansion~\eqref{V-expansion} are completely determined by $u_i$:
\begin{equation}\label{A-matrix}
    A_{ij}^{\xi} = 
    \begin{cases}
    r_{\xi}^2 + \sum_{k=1, k \neq j}^n a_\xi / |u_j - u_k|^3,
    \quad i = j,\\
    -a_{\xi}/|u_i - u_j|^3,
    \quad i \neq j,
    \end{cases}
\end{equation}
where $r_{\xi} = \omega_\xi/\omega_z$, $a_x = a_y = -1$ and $a_z = 2$. The eigen frequencies $\omega_{\xi,m} = \sqrt{\lambda_{\xi, m}} \, \omega_{\xi}$ and eigen vectors $\bm{b}^{\xi,m}$ for both the radial ($\xi = x$) and axial ($\xi = z$) normal modes $m$ are obtained from diagonalization of the matrix~\eqref{A-matrix},
\begin{equation}\label{A-eigen}
    A^{\xi} \bm{b}^{\xi,m} 
    = \lambda_{\xi,m}\, \bm{b}^{\xi,m},
    \quad
    m=1,\ldots,n.
\end{equation}
Note that the normal modes along $x$ and $y$ directions are degenerate, so it is enough to consider only the $x$-modes. 

We diagonalized the matrix \eqref{A-matrix} to calculate its phonon eigen frequencies $\omega_{m} \equiv \sqrt{\lambda_{x,m}} \, \omega_x$ and eigen vectors $\bm{b}^m \equiv \bm{b}^{x,m}$ for the radial normal modes ($\xi = x$). The contribution from the axial normal modes ($\xi = z$) was not taken into account due to the geometry of chosen laser beams. The Lamb-Dicke parameters were calculated then as
\begin{equation}\label{LD-params-1}
    \eta_i^{m} = \delta k \sqrt{\hbar/(2 M \omega_{m})} \, b_i^{m}.
\end{equation}
Here, $\delta k = 2\pi/\lambda$, where $\lambda$ is the wavelength of a laser beam. After that, we substituted the frequencies $\omega_{m}$ and Lamb-Dicke parameters $\eta_i^{m}$ into Eq.~\eqref{J-def} to calculate the matrix $C_{ij}$ for the Ising couplings in the ion Hamiltonian for a given laser detuning $\mu$.

\section{\label{appendix:numerics}Numerical details}

For each number of ions (qubits) $n$, we simulated a chain of $^{40}$Ca$^+$ ions of the atomic mass $M=39.96$ a.m.u. The radial and axial frequencies for a linear trap were set to $\omega_x/2\pi = 1$ MHz and $\omega_z/2\pi = 0.15$ MHz, respectively. The wavelength of a laser beam was set to $\lambda = 729.15$ nm. After calculating the phonon frequencies $\omega_m$ and Lamb-Dicke parameters $\eta_i^m$ as described in Appendix~\ref{appendix:phonons}, we evaluated the matrix $C_{ij}$ using Eq.~\eqref{J-def} with the laser detuning fixed at $\mu = \omega_x + 2 \pi \delta \nu$, where $\delta \nu = 1$ kHz. The maximal Rabi frequency in Eq.~\eqref{J-modified} was chosen equal to $\Omega_{\max}/2\pi = 30$ kHz for all $n$ (except for 15 qubits, where we used 50 kHz). This choice of $\Omega_{\max}$ allowed to normalize the matrix~\eqref{J-modified} of Ising couplings such that its maximal element was about $J_{\max} \sim 1$ kHz. The latter implies propagation times with the ion Hamiltonian $H_{\rm I}$ of the order of several milliseconds~\cite{rabinovich2022ion}.

For each sampled instance of the SK model~\eqref{H-SK}, we run the heuristic proposed in Sec.~\ref{sec:heuristics} to obtain a problem-specific configuration $\bm{A}$ of the ion native ansatz by minimizing the energy~\eqref{cost-energy-single-layer} for a single layer alternately over variational and hyper- parameters. On each BCD iteration~$k$, we minimized the energy $F(\bm{\theta}) = E(\bm{\theta},\bm{A}_k)$ over the block of variational parameters $\bm{\theta} = (\beta, \gamma) \in \Theta$ using the approach inspired by~\cite{PPY2024}.
We evaluated the energy $F(\bm{\theta})$ on a $10 \times 10$ discretization grid introduced on $\Theta$ and stored these values in the form of a matrix $F_{ij} = F(\beta_i, \gamma_j)$. Then, we found the position of its minimal element, $(i_*,j_*) = \argmin_{i,j} F_{ij}$, and used $(\beta_{i_*},\gamma_{j_*})$ as a starting point for the standard gradient-based Broyden-Fletcher-Goldfarb-Shanno (BFGS)~\cite{DS96} optimizer, which locally converged to the optimal solution~$\bm{\theta}_k$. This technique allows to overcome optimization issues and avoid local minima. 
For minimizing the energy $E(\bm{\theta}_k,\bm{A})$ over the block of hyperparameters $\bm{A} \in [-1,1]^n$,
we used the efficient derivative-free Powell's conjugate directions method~\cite{Powell1964}. Both optimizers, L-BFGS-B and Powell's, were taken as implemented in the SciPy library~\cite{SciPy} using default settings. The parameters of the BCD loop were set equal to $k_{\max} = 50$, $m_{\max}=10$ and $\delta = 10^{-3}$.

For each sampled SK instance, we searched the optimal scaling factor $\alpha^*$ for obtained hyperparameters $\bm{A}^*$ to maximize the fraction of parameters space below some certain value. The latter was estimated by the cost function $f(\alpha)$ introduced at Algorithm~\ref{alg:rescaling}. The constructed cost function $f(\alpha)$ was maximized by means of the COBYLA optimizer~\cite{Powell1994} taken from~\cite{SciPy}. Only a few iterations of the optimizer starting from $\alpha_0=0.8$ were required to reach the maximum with a satisfactory accuracy. The grid size for evaluating the cost landscape~\eqref{cost-energy-single-layer} was taken equal to $N=20$. The parameters were set equal to $\epsilon = 0.05$ and $\mu=0.95$.

The optimization of the QAOA energy \eqref{cost-energy} was performed by means of a layerwise inspired heuristic~\cite{akshay2022circuit}. This method iterates over the QAOA circuit depth $p$ starting from a single layer ($p=1$) that is optimized with respect to two variational parameters. On each iteration, a new layer is added and optimized separately over its variational parameters, while keeping all other parameters from the previous layers fixed. This step is followed by the simultaneous optimization over all variational parameters for the whole QAOA circuit. For optimization, we once again used the L-BFGS-B optimizer. On each step of this layerwise training heuristic, we started the optimizer from 25 random initial points sampled uniformly in the parameter space and chose the best result. In order to find a global minimum, we performed 10 runs of this layerwise training for each circuit depth $p$ in our QAOA simulations choosing the result with the smallest energy~\eqref{cost-energy}. 

\section{\label{appendix:stateprep}State preparation problem}
Minimization of a diagonal problem Hamiltonian requires one to find an approximation to a bit string state $\ket{t}$ of the lowest energy. Thus, to estimate the depth sufficient to solve an arbitrary combinatorial problem, one only needs to find the depth of a circuit that can prepare an arbitrary bit string state. For the case of a $\mathbb{Z}_2$-symmetric ansatz and problem Hamiltonian, the latter translates to preparing a GHZ-like states of the form $\dfrac{1}{\sqrt{2}}(\ket{t}+X^{\otimes n}\ket{t})$. 

Importantly, the ansatz \eqref{ion-native-ansatz} possesses certain symmetries with respect to hyperparameters. Specifically, given a state $\ket{\psi(\bm\beta,\bm\gamma,A_j)}$, prepared using ansatz \eqref{ion-native-ansatz} with hyperparameters $\{A_j\}$, one can note
\begin{eqnarray}
\label{A_flip}
    X_m \ket{\psi(\bm\beta,\bm\gamma,A_j)}=\ket{\psi(\bm\beta,\bm\gamma, (-1)^{\delta_{jm}}A_j)}. 
\end{eqnarray}
In other words, flipping the $m$-th bit in the prepared state is equivalent to flipping the sign of $A_m$. By extension, flipping several qubits is equivalent to changing the signs of the corresponding hyperparameters.
As a consequence, the same circuit depth is required to prepare any bit string $\ket{t}$ --- or rather its symmetrized version $\dfrac{1}{\sqrt{2}}\big(\ket{t}+X^{\otimes n}\ket{t}\big)$, which respects the $\mathbb{Z}_2$ symmetry of the ansatz ~\cite{rabinovich2022ion}. Thus, the minimization of all diagonal Hamiltonians of a given size requires the same circuit depth, at least in principle. This argument evidently ignores how easy it is to train such circuits as well as how to properly adjust the ansatz hyperparameters. Nevertheless, this fact can be used to establish a bound on the required depth to solve specific problems.

Due to property \eqref{A_flip} of the ion native ansatz, one can restrict the problem to the preparation of any of these states. 
Therefore, here we maximize the overlap with the GHZ state,
\begin{equation}\label{app-state-prep}
    |\braket{\psi_p(\bm{\beta},\bm{\gamma})}{\rm GHZ}|^2 \to \max_{\bm{\beta},\bm{\gamma}},
\end{equation}
where $\ket{\rm GHZ} = (\ket{0}^{\otimes n} + \ket{1}^{\otimes n})/\sqrt{2}$.

\begin{figure}[ht!]
    \centering
    \includegraphics[width=0.9\linewidth]{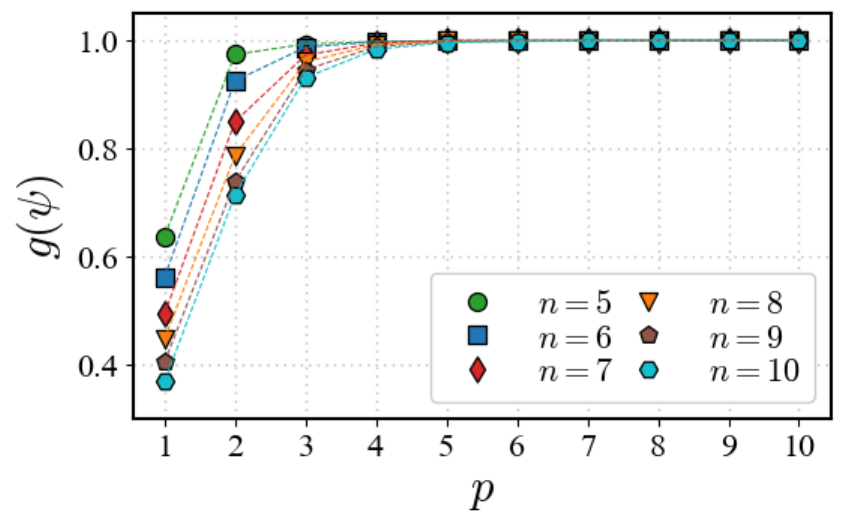}
    \caption{Ground state overlap $g(\psi)$ as a function of circuit depth $p$ calculated by the ion native QAOA when solving the state preparation problem~\eqref{app-state-prep} for different system sizes~$n$. For each $n$, the results are obtained using problem-specific hyperparameters found by the proposed heuristic.} 
    \label{fig:stateprep}
\end{figure}

Following our numerics for the SK Hamiltonian~\eqref{H-SK} in Sec.~\ref{sec:results}, we considered system sizes $n$ of $5$--$10$ qubits. We use the proposed heuristic, by minimizing the Hamiltonian $H_{\rm P} = -(\ketbra{0}{0}^{\otimes n} + \ketbra{1}{1}^{\otimes n})$ which, due to the $\mathbb{Z}_2$ symmetry, is equivalent to the state preparation problem~\eqref{app-state-prep}. The heuristic was tuned by setting $\varepsilon = 0$ in the convergence criteria~\eqref{heuristics-criteria} to identify ansatz hyperparameters providing the smallest energy~\eqref{cost-energy-single-layer} (largest overlap \eqref{app-state-prep}) over $m_{\max}=10$ runs. In this setting, the heuristic tried to reach a global minimum of the energy for a single layer.
The rescaling parameter was set to $\mu=1.1$.
Once the ansatz hyperparameters were identified, we performed ion native QAOA simulations for circuit depths $p$ of up to 10 layers. 
Fig.~\ref{fig:stateprep} shows the ground state overlap $g(\psi)$ calculated as a function of the QAOA circuit depth $p$. 
As one can see, the QAOA performance slightly decreases with the system size $n$. Nevertheless, $p=2$ layers are already sufficient to reach the ground state overlap of $50\%$ for all considered $n$, while $p=7$ layers guarantee the overlap of $99.9\%$.

\end{document}